\documentclass[a4paper]{article}
\usepackage{pdfsync}
\usepackage[utf8]{inputenc}
\usepackage{graphicx}
\usepackage{amssymb,amsfonts,amsmath,amsthm}
\usepackage{hyperref}
\usepackage{graphicx}
\usepackage{multirow}
\usepackage{verbatim}
\usepackage{color}
\usepackage{booktabs}
\usepackage[sectionbib,round]{natbib}
\usepackage{textcomp}
\usepackage{upquote}
\usepackage{bm}
\usepackage{dirtree}

\theoremstyle{plain}

\theoremstyle{definition}

\begin{document}

\title{\textbf{Conventional and Fuzzy Data Envelopment Analysis with deaR}}


\author{V. J. Bol\'os$^1$, R. Ben\'{\i}tez$^1$, V. Coll-Serrano$^2$ \\ \\
{\small $^1$ Dpto. Matem\'aticas para la Econom\'{\i}a y la Empresa, Facultad de Econom\'{\i}a.} \\
{\small $^2$ Dpto. Econom\'{\i}a Aplicada, Facultad de Econom\'{\i}a.} \\
{\small Universidad de Valencia. Avda. Tarongers s/n, 46022 Valencia, Spain.} \\
{\small e-mail\textup{: \texttt{vicente.bolos@uv.es}, \texttt{rabesua@uv.es}, \texttt{vicente.coll@uv.es}}} \\}

\date{November 2022}

\maketitle

\begin{abstract}
\textbf{deaR} is a recently developed \textsf{R} package for data envelopment analysis (DEA) that implements a large number of conventional and fuzzy models, along with super-efficiency models, cross-efficiency analysis, Malmquist index, bootstrapping, and metafrontier analysis. It should be noted that \textbf{deaR} is the only package to date that incorporates Kao-Liu, Guo-Tanaka and possibilistic fuzzy models. The versatility of the package allows the user to work with different returns to scale and orientations, as well as to consider special features, namely non-controllable, non-discretionary or undesirable variables. Moreover, it includes novel graphical representations that can help the user to display the results. This paper is a comprehensive description of \textbf{deaR}, reviewing all implemented models and giving examples of use.
\end{abstract}

\section{Introduction}

Data envelopment analysis (DEA) \citep{Charnes1978} is a non-parametric technique used to measure the relative efficiency of a homogeneous set of decision making units (DMUs) that use multiple inputs to obtain multiple outputs. Using mathematical programming methods, DEA allows the identification of the best practice frontier (efficient frontier). DMUs that form the best practice frontier are qualified as efficient, while DMUs that move away from the frontier are inefficient.

DEA has been used to evaluate efficiency in many different fields such as education, agriculture and farm, banking, health, transportation, public administration, etc. Recently, \citet{Emrouznejad2018} have compiled a list of more than $10000$ articles related to DEA. Along with the methodological advances and development of new DEA models and practical applications, software (commercial and non-commercial) has also appeared to facilitate its use by both researchers and practitioners. \citet{Daraio2019} perform a review of DEA software available to assess efficiency and productivity and compare their different options. Among the commercial software, MaxDEA (\url{http://maxdea.com/MaxDEA.htm}), DEA-Solver-PRO \citep{Cooper2007} and DEAFrontier \citep{Zhu2014} stand out for the diversity of DEA models implemented. Also, some of the most popular software, such as \textsf{SAS}, \textsf{GAMS} or \textsf{Stata}, include modules to estimate efficiency using basic DEA models. Moreover, \citet{Alvarez2020} have written a package for \textsf{MATLAB} that includes several DEA models like radial DEA, directional distance function, Malmquist index and cross-efficiency among others.

As far as non-commercial software is concerned, one of the most widely used is undoubtedly DEAP, developed by \citet{Coelli1996}. However, this software is very limited in terms of the variety of DEA models it can solve: basic models by \citet{Charnes1978} and \citet{Banker1984} (referred to as CCR and BCC models), Malmquist index (FGNZ decomposition due to \citet{Fare1994}) and cost/revenue models.

Nowadays, the use of non-commercial software to apply DEA models involves the use of \textsf{R} packages.
The first \textsf{R} packages for measuring efficiency and productivity using DEA were \textbf{FEAR} by \citet{Wilson2008} (it is distributed under license and has recently been updated after many years without doing so) and \textbf{Benchmarking} by \citet{Bogetoft2022}, which complements the book by the same authors \citep{Bogetoft2011}. These two packages contain methods that allow the use of different technology assumptions and different efficiency measures (radial measures, super-efficiency, additive models, cost efficiency, etc.), but it is worth noting that they include routines to apply the bootstrap methods described by \citet{Simar1998}. Without being exhaustive, other relevant packages available in \textsf{R} for analysing efficiency and productivity using the DEA technique are:

\begin{itemize}
\item \textbf{nonparaeff} \citep{Oh2013} includes functions to solve, among others, the following models: CCR, BCC, additive, SBM, assurance region, cost and revenue, the FGNZ decomposition of the Malmquist index, and the directional distance function with undesirable outputs. The latest update is February 2013. 

\item \textbf{DJL} \citep{Lim2022} allows to apply the basic DEA models but above all, we can highlight that this package includes functions to apply network DEA \citep{Cook2010} and dynamic DEA \citep{Kao2013,Emrouznejad2005}.

\item \textbf{additiveDEA} \citep{Soteriades2017} uses two types of additive models to calculate efficiency. The user can choose the SBM model (Tone, 2001) or generalized additive models: range adjusted measure (RAM) \citep{Cooper1999,Cooper2001}, bounded adjusted measure (BAM) \citep{Cooper2011}, measure of inefficiency proportions (MIP) \citep{Cooper1999}, or the Lovell-Pastor measure (LovPast) \citep{Knox1995}. This package has not been updated since October 2017.

\item \textbf{rDEA} \citep{Simm2020}. With this package the user can apply the CCR, BCC or cost-minimization models and, as a differential aspect, estimate robust efficiency scores with or without exogenous variables. For this purpose, the corresponding functions implement the algorithms of \citet{Simar1998,Simar2007}. It was last updated in early 2020.
\end{itemize}

Unlike the \textsf{R} packages cited above, \textbf{deaR} \citep{Coll2022} has a wide variety of models implemented that allow the user to apply both conventional and fuzzy models, which consider imprecision or uncertainty in the data. Moreover, the package includes super-efficiency, cross-efficiency, Malmquist index and bootstrapping models. The functions in \textbf{deaR} are designed to be easily used by a non-expert \textsf{R} user. 

This paper is organized as follows. In Section \ref{sec:deaR} we describe how to use the \textbf{deaR} package in 3 steps: introducing data, running a model and extracting the results. We also present the main \textsf{S3} data classes and review some basic radial efficiency models. In Section \ref{sec:models} we review other conventional efficiency models such as multiplier, free disposal hull, directional, non-radial, additive or SBM models. In Section \ref{secund} we introduce some special features on data, namely non-controllable, non-discretionary and undesirable variables. Section \ref{sec:super} is devoted to the main super-efficiency models: radial, SBM and additive models. In Section \ref{sec:cross} we present the cross-efficiency analysis, while in Section \ref{sec:fuzzy} we review some popular fuzzy models: Kao-Liu, Guo-Tanaka and possibilistic models. We show the Malmquist productivity indices in Section \ref{sec:malm}, and the bootstrapping methodology in Section \ref{sec:boots}. Finally, we give an example on how to perform a non-parametric metafrontier analysis in Section \ref{sec:metaf} and provide some conclusions in Section \ref{sec:conc}.

\section[The deaR package]{The \textbf{deaR} package}
\label{sec:deaR}

Throughout this paper, we consider $\mathcal{D}=\left\{ \textrm{DMU}_1, \ldots ,\textrm{DMU}_n \right\} $ a set of $n$ DMUs with $m$ inputs and $s$ outputs. Matrices $X=(x_{ij})$ and $Y=(y_{rj})$ are the \emph{input} and \emph{output data matrices}, respectively, where $x_{ij}$ and $y_{rj}$ denote the $i$-th input and $r$-th output of the $j$-th DMU. We also assume that $x_{ij}$ and $y_{rj}$ are all positive, i.e., greater than $0$. Nevertheless, this restriction is not strictly considered in \textbf{deaR} package, and we can run models with negative or zero data, but results should be interpreted with care. In general, we denote vectors by bold-face letters and they are considered as column vectors unless otherwise stated. The elements of a vector are denoted by the same letter as the vector, but unbolded and with subscripts. The $0$-vector is denoted by $\bm{0}$ and the context determines its dimension.

\subsection[What does deaR do?]{What does \textbf{deaR} do?}

First of all, \textbf{deaR} is a package available on CRAN and it can be installed with
\begin{verbatim}
R> install.packages("deaR")
\end{verbatim}

In order to perform a DEA study over DMUs, we need to solve optimization problems. Specifically, all the models implemented in \textbf{deaR} involve linear programming problems. 
There are many options to solve linear programs using \textsf{R}, as shown in the Optimization Task View (linear programming section) but, in \textbf{deaR}, we use the \textbf{lpSolve} package \citep{Berkelaar2020}.
In fact, \textbf{deaR} can be considered as a wrapper of \textbf{lpSolve}.

\begin{figure}[htbp]
  \centering
  \includegraphics[width = .8\linewidth]{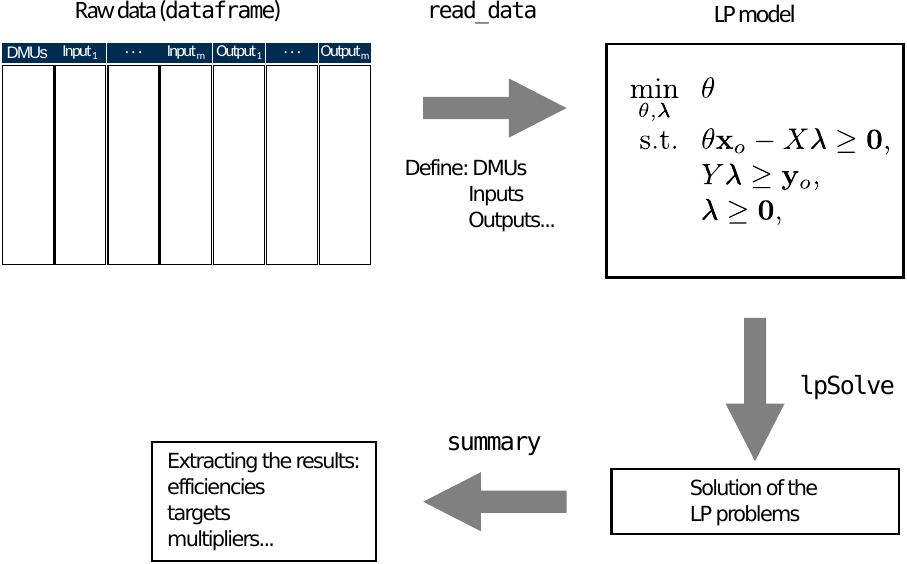}
  \caption{\textbf{deaR} workflow. From raw data, we construct the corresponding linear programming models which are solved with \textbf{lpSolve}. Finally, parameters of interest are extracted from the solution.}
  \label{fig:figure1}
\end{figure}

The workflow in \textbf{deaR} has the following steps (see Figure~\ref{fig:figure1}):
\begin{enumerate}
\item From raw data, parameters of the problem (DMUs, inputs, outputs, etc.) are defined.
\item The corresponding linear programming models (objective functions and constraints) are built.
\item The linear programs are solved with \textbf{lpSolve}.
\item Different parameters of interest are extracted from the solution (efficiency scores, targets, multipliers, etc.).
\end{enumerate}

Therefore, performing a DEA analysis with \textbf{deaR} can be divided into $3$ phases: introducing data, running a model and extracting the results. We will describe each phase in detail below.

\subsection[Introducing data: The deadata class]{Introducing data: The \texttt{deadata} class}

Datasets in DEA are usually formatted in a spreadsheet table, in which one of the columns (typically the first one) corresponds to the DMUs identification labels, the next $m$ columns correspond to the different inputs, and the following $s$ columns correspond to the outputs. This type of data arrangement will be referred to as \emph{standard} DEA dataset.

The package contains more than 20 datasets from different books and research papers, that can be used to reproduce the results obtained in those sources (use \texttt{data(package = "deaR")} to get the list of all available datasets). For example, the \texttt{Fortune500} dataset is a standard DEA dataset containing 15 companies from the 1995's Fortune 500 list \citep{Zhu2014}:

\begin{verbatim}
R> library("deaR")
R> head(Fortune500)
\end{verbatim}
\begin{verbatim}
         Company   Assets  Equity Employees  Revenue Profit
1     Mitsubishi  91920.6 10950.0     36000 184365.2  346.2
2         Mitsui  68770.9  5553.9     80000 181518.7  314.8
3         Itochu  65708.9  4271.1      7182 169164.6  121.2
4 General Motors 217123.4 23345.5    709000 168828.6 6880.7
5       Sumitomo  50268.9  6681.0      6193 167530.7  210.5
6       Marubeni  71439.3  5239.1      6702 161057.4  156.6
\end{verbatim}

In this standard DEA dataset, the first column (``Company'') contains the names of the different firms included. The next three columns (``Assets'', ``Equity'' and ``Employees'') define the inputs, and the last two columns (``Revenue'' and ``Profit'') correspond to the outputs. 

The package \textbf{deaR} first needs to read the dataset and identify the DMUs, inputs, outputs and all other relevant information about the data. To do so, we will make use of the function \texttt{make\_deadata}. For a standard DEA dataset, the syntax is straightforward. By default, the function assumes that the names of the DMUs are in the first column (although this can be changed with the parameter \texttt{dmus}) and therefore, we will only need to declare the number of inputs (\texttt{ni}) and the number of outputs (\texttt{no}):

\begin{verbatim}
R> dataFortune <- make_deadata(Fortune500, ni = 3, no = 2)
\end{verbatim}

Instead of defining the number of inputs/outputs with the parameters \texttt{ni/no}, we can explicitly define which columns contain the inputs and which the outputs, either by a number (position of the column) or by a column name. For instance, we can define a smaller DEA dataset with two inputs and one output without having to subset the dataframe:

\begin{verbatim}
R> dataFortuneSmall <- make_deadata(Fortune500, inputs = c(3, 4),
+                                   outputs = "Profit")
\end{verbatim}
In this example, the DEA dataset have ``Equity'' and ``Employees'' as inputs (columns 3 and 4) and ``Profit'' as output (defined by the column name). Alternatively, we can explicitly define the input and output data matrices, with DMUs in columns:

\begin{verbatim}
R> inputs <- matrix(c(10950, 5553.9, 4271.1, 36000, 80000, 7182),
+                   nrow = 2, ncol = 3, byrow = TRUE,
+                   dimnames = list(c("Equity", "Employees"),
+                                   c("Mitsubishi", "Mitsui", "Itochu")))
R> outputs <- matrix(c(346.2, 314.8, 121.2),
+                    nrow = 1, ncol = 3, byrow = TRUE,
+                    dimnames = list("Profit",
+                                    c("Mitsubishi", "Mitsui", "Itochu")))
R> dataFortuneSmall2 <- make_deadata(inputs = inputs, outputs = outputs)
\end{verbatim}
In this case, we only consider ``Equity'', ``Employees'' and ``Profit'' of the first three DMUs. Moreover, if names are not provided by \texttt{dimnames}, then they are automatically generated as ``DMU1'', ``DMU2'', ``Input1'', ``Input2'', etc.

It is important to note that negative or zero data are not allowed by some models. To solve this problem, it is recommended to translate the base point of the inputs/outputs with negative or zero data in order to get only positive values. Nevertheless, depending on the nature of the data and the model, this may not be appropriate in some cases.
Moreover, if there are data with very different orders of magnitude, then it is also recommended to redefine the units of measure in order to prevent ill-posed linear problems.

Finally, the resulting value of the \texttt{make\_deadata} function is an object of class \texttt{deadata} which is a list with the following fields: 
\dirtree{%
.1 deadata (\textrm{for a model with $n$ DMUs, $m$ inputs and $s$ outputs}.).
.2 input: \textrm{input data matrix of size $m\times n$}..
.2 output: \textrm{output data matrix of size $s\times n$}..
.2 dmunames: \textrm{character vector containing the names of the DMUs}..
.2 nc\_inputs: \textrm{integer vector identifying the non-controllable inputs (or \texttt{NULL})}..
.2 nc\_outputs: \textrm{integer vector identifying the non-controllable outputs (or \texttt{NULL})}..
.2 nd\_inputs: \textrm{integer vector identifying the non-discretionary inputs (or \texttt{NULL})}..
.2 nd\_outputs: \textrm{integer vector identifying the non-discretionary outputs (or \texttt{NULL})}..
.2 ud\_inputs: \textrm{integer vector identifying the undesirable inputs (or \texttt{NULL})}..
.2 ud\_outputs: \textrm{integer vector identifying the undesirable outputs (or \texttt{NULL})}..
}
Special features like non-controllable, non-discretionary and undesirable inputs/outputs are explained in Section \ref{secund}.

\subsection{Running a model}
\label{sec:run}

In general, a DMU is \emph{efficient} if there is not any feasible activity in a given production possibility set ``better than'' the DMU, in the sense that consumes less inputs and produces more outputs. Hence, an \emph{efficiency model} first establishes the production possibility set and then it checks if the DMU is efficient or not. If the DMU is inefficient, then the model usually gives a score and a target for improving the activity.

Once the data has been read and we have an object of class \texttt{deadata}, we can proceed to select and run a model. The package \textbf{deaR} has quite a wide range of different models available. Table~\ref{tab:models} lists all the model functions included in \textbf{deaR}. In this section we shall illustrate the use of those functions with classical examples of the basic radial CCR and BCC models contained in function \texttt{model\_basic}.

\begin{table}
\small
\label{tab:models}
\centering
\begin{tabular}{p{.28\textwidth}p{.33\textwidth}p{.3\textwidth}}
\toprule
Function name& Models description& References\\\midrule
\texttt{model\_additive} & Additive models. & \text{}\cite{Charnes1985}.\\
\texttt{model\_addmin} & Additive-Min models. & \text{}\cite{Aparicio2007}.\\
\texttt{model\_addsupereff}& Additive super-efficiency models. &\text{}\cite{Du2010}.\\
\texttt{model\_basic}& Basic radial models, such as CCR and BCC, and directional models. &\text{}\cite{Charnes1978,Charnes1979}; 

\cite{Banker1984,Chambers1996,Chambers1998}.\\
\texttt{model\_deaps}& Non-radial DEA preference structure models. &\text{}\cite{Zhu1996}.\\
\texttt{model\_fdh}& Free disposal hull models. &\text{}\cite{Thrall1999}.\\
\texttt{model\_multiplier}& Radial models, such as CCR and BCC models, in multiplier form. &\text{}\cite{Charnes1962}.\\
\texttt{model\_nonradial}& Non-radial models. &\text{}\cite{Fare1978,Wu2011}.\\
\texttt{model\_profit}& Cost, revenue and profit ef\-fi\-cien\-cy DEA models. &\text{}\cite{Coelli2005}.\\
\texttt{model\_rdm}& Range directional models. &\text{}\cite{Portela2004}.\\
\texttt{model\_sbmeff}& Slacks-based measure of ef\-fi\-cien\-cy models. &\text{}\cite{Tone2001}.\\
\texttt{model\_sbmsupereff}& Slacks-based measure of super-ef\-fi\-cien\-cy models. &\text{}\cite{Tone2002,Tone2010}.\\
\texttt{model\_supereff}& Radial super-efficiency models. &\text{}\cite{Andersen1993}.\\ \midrule
\texttt{modelfuzzy\_kaoliu}& Kao-Liu fuzzy models. &\text{}\cite{Kao2000a,Kao2000b,Kao2003}.\\
\texttt{modelfuzzy\_guotanaka}& Guo-Tanaka fuzzy models. &\text{}\cite{Guo2001}.\\
\texttt{modelfuzzy\_possibilistic}& Possibilistic fuzzy models. &\text{}\cite{Leon2003}.\\ \midrule
\texttt{cross\_efficiency}& Arbitrary, benevolent and aggressive cross-efficiency.  &\text{}\cite{Doyle1994,Cook2015,Lim2015a}.\\
\texttt{cross\_efficiency\_fuzzy}& Cross-efficiency analysis from a Guo-Tanaka model solution. &\text{}\cite{Doyle1994,Guo2001}.\\ \midrule
\texttt{malmquist\_index}& Malmquist productivity index for productivity change over time. &\text{}\cite{Fare1997,Fare1998}.\\ \midrule
\texttt{bootstrap\_basic}& Bootstrap efficiency scores. &\text{}\cite{Simar1998}. \\
\bottomrule
\end{tabular}
\caption{DEA models available in package \textbf{deaR}.}
\end{table}


The first model we are going to introduce is the so called CCR model \citep{Charnes1978,Charnes1979,Charnes1981}. This model assumes that the production possibility set, i.e., the set of feasible activities defined by the set $\mathcal{D}$ of DMUs, is under constant returns to scale (CRS) and given by
\begin{equation}
\label{eq:p}
P =P(X,Y)=\left\{ \left( \mathbf{x},\mathbf{y} \right) \in \mathbb{R}_{>0}^{m+s}\ \ | \ \ \mathbf{x}\geq X\bm{\lambda},\quad \mathbf{y}\leq Y\bm{\lambda},\quad \bm{\lambda}\geq \mathbf{0} \right\} ,
\end{equation}
where $X$ and $Y$ are the input and output data matrices, respectively, and $\bm{\lambda}=(\lambda_1,\ldots,\lambda_n)^\top$ is a column vector.
Hence, $\textrm{DMU}_o\in\mathcal{D}$ is efficient if and only if there is no $(\mathbf{x},\mathbf{y})\in P$ such that $x_{io}\geq x_i$ and $y_{ro}\leq y_r$ with at least one strict inequality.

The CCR model can be either input or output-oriented. In the former case (see \eqref{eq:ccr} (a)) we look for determining the maximal proportionate reduction of inputs allowed by the production possibility set, while maintaining the current output level of $\textrm{DMU}_o$. On the other hand, in the output-oriented case (see \eqref{eq:ccr} (b)), we want to find the maximal proportionate increase of outputs while keeping the current input consumption of $\textrm{DMU}_o$:

\begin{equation}
\label{eq:ccr}
\def\arraystretch{1.2}
\begin{array}{ll}
\begin{array}[t]{rl}
\textrm{(a) }\min \limits_{\theta, \bm{\lambda}} & \theta \\
\text{s.t.} & \theta \mathbf{x}_o - X\bm{\lambda}\geq \mathbf{0}, \\
& Y\bm{\lambda} \geq \mathbf{y}_o, \\
& \bm{\lambda}\geq \mathbf{0},
\end{array}
&\qquad \qquad
\begin{array}[t]{rl}
\textrm{(b) }\max \limits_{\eta, \bm{\lambda}} & \eta \\
\text{s.t.} & X\bm{\lambda}\leq \mathbf{x}_o, \\
& \eta \mathbf{y}_o-Y\bm{\lambda} \leq \mathbf{0}, \\
& \bm{\lambda}\geq \mathbf{0},
\end{array}
\end{array}
\end{equation}
where $\mathbf{x}_o=(x_{1o},\ldots,x_{mo})^\top$ and $\mathbf{y}_o=(y_{1o},\ldots,y_{so})^\top$ are column vectors.
In a second stage, with our knowledge of the optimal objectives $\theta ^*$ or $\eta ^*$, we solve the following linear program, \eqref{eq:ccr2} (a) or \eqref{eq:ccr2} (b), in order to find the \emph{max-slack solution}:
\begin{equation}
\label{eq:ccr2}
\def\arraystretch{1.2}
\begin{array}{ll}
\begin{array}[t]{rl}
\textrm{(a) }\max \limits_{\bm{\lambda},\mathbf{s}^-,\mathbf{s}^+} &\omega = \mathbf{w}^-\mathbf{s}^-+\mathbf{w}^+\mathbf{s}^+ \\
\text{s.t.} & X\bm{\lambda}+\mathbf{s}^-=\theta ^*\mathbf{x}_o, \\
& Y\bm{\lambda} -\mathbf{s}^+=\mathbf{y}_o, \\
& \bm{\lambda}\geq \mathbf{0},\,\, \mathbf{s}^-\geq \mathbf{0},\,\, \mathbf{s}^+\geq \mathbf{0},
\end{array}
&\qquad
\begin{array}[t]{rl}
\textrm{(b) }\max \limits_{\bm{\lambda},\mathbf{s}^-,\mathbf{s}^+} &\omega = \mathbf{w}^-\mathbf{s}^-+\mathbf{w}^+\mathbf{s}^+ \\
\text{s.t.} & X\bm{\lambda}+\mathbf{s}^-=\mathbf{x}_o, \\
& Y\bm{\lambda} -\mathbf{s}^+=\eta ^*\mathbf{y}_o, \\
& \bm{\lambda}\geq \mathbf{0},\,\, \mathbf{s}^-\geq \mathbf{0},\,\, \mathbf{s}^+\geq \mathbf{0},
\end{array}
\end{array}
\end{equation}
where the weights $\mathbf{w}^-$ and $\mathbf{w}^+$ are positive row vectors.
In the input-oriented CCR model, $\textrm{DMU}_o$ is efficient if and only if $\theta ^*=1$ and $\omega ^*=0$. If $\textrm{DMU}_o$ is inefficient, then $0<\theta ^*\leq 1$ is the \emph{efficiency score} and $(X\bm{\lambda}^*,Y\bm{\lambda}^*)$ is the \emph{target}, that can be interpreted as the projection of $\textrm{DMU}_o$ onto the \emph{efficient frontier}. Note that there can be inefficient DMUs with $\theta ^*=1$, called \textit{weakly efficient}. On the other hand, in the output-oriented CCR model, we have $\eta ^*=1/\theta ^*$.

The BCC model \citep{Banker1984} considers the production possibility set under variable returns to scale (VRS),
\begin{equation}
\label{eq:pb}
P_B =P_B(X,Y)=\left\{ \left( \mathbf{x},\mathbf{y} \right) \in \mathbb{R}_{>0}^{m+s}\ \ | \ \ \mathbf{x}\geq X\bm{\lambda},\quad \mathbf{y}\leq Y\bm{\lambda},\quad \mathbf{e}\bm{\lambda} =1,\quad \bm{\lambda}\geq \mathbf{0} \right\} ,
\end{equation}
where $\mathbf{e}=(1,\ldots ,1)$ is a row vector. Oriented BCC models are constructed by adding $\mathbf{e}\bm{\lambda} =1$ to the constraints of \eqref{eq:ccr} and \eqref{eq:ccr2}. Efficiency scores and targets are defined analogously to the CCR model.

In the production possibility set, the returns to scale condition can be changed to non-increasing (NIRS), non-decreasing (NDRS) or generalized returns to scale (GRS). In these cases, the condition $\mathbf{e}\bm{\lambda}=1$ is replaced by $0\leq \mathbf{e}\bm{\lambda}\leq 1$ (NIRS), $\mathbf{e}\bm{\lambda}\geq 1$ (NDRS) or $L\leq \mathbf{e}\bm{\lambda}\leq U$ (GRS), with $0\leq L\leq 1$ and $U\geq 1$.
These conditions are added to the constraints of \eqref{eq:ccr} and \eqref{eq:ccr2} in order to build models with different returns to scale.

The syntax of \texttt{model\_basic} is very flexible, and contains a great deal of parameters allowing the user to run different models from within the same function. The main parameters are:
\begin{itemize}
\item \texttt{datadea}: an object of class \texttt{deadata} (e.g., the output of \texttt{make\_deadata} function).
\item \texttt{orientation}: orientation of the model. It can be either input-oriented (\texttt{"io"}, by default), output-oriented (\texttt{"oo"}) or directional (\texttt{"dir"}), as we will see in Section \ref{secdir}.
\item \texttt{rts}: returns to scale regime of the model. It can be either \texttt{"crs"} (by default), \texttt{"vrs"}, \texttt{"nirs"}, \texttt{"ndrs"} or \texttt{"grs"}. If the \texttt{"grs"} option is selected, then the two optional parameters \texttt{L} and \texttt{U} should be given. By default, $\text{\texttt{L}}=\text{\texttt{U}}=1$.
\end{itemize}

Other optional parameters of interest are:
\begin{itemize}

\item \texttt{dmu\_eval} and \texttt{dmu\_ref}: Those are numeric vectors. The former determines which DMUs are going to be evaluated while the latter defines the \emph{evaluation reference set}, i.e., with respect to which DMUs we are going to evaluate. Note that the production possibility set is constructed taking into account only the DMUs in \texttt{dmu\_ref}. If \texttt{dmu\_eval} or \texttt{dmu\_ref} are not provided by the user, then all the DMUs are considered. These parameters are used, for example, for conducting a non-parametric metafrontier analysis when the DMUs set is divided into several groups, as we will see in Section \ref{sec:metaf}.

\item \texttt{maxslack}: If this logical variable is set to \texttt{TRUE} (by default), then the max-slack solution is computed in a second stage (see \eqref{eq:ccr2} for the CCR model). Weights  $\mathbf{w}^-$ and $\mathbf{w}^+$ for each DMU are defined with the \texttt{weight\_slack\_i} and \texttt{weight\_slack\_o} parameters respectively, which can be either a vector of weights (one for each input/output), or even a matrix of size [number of inputs/outputs]$\times$[number of DMUs in \texttt{dmu\_eval}]. Then, not only each input/output may have a different weight, but also they can change with the DMUs.

\item \texttt{returnlp}: If this logical variable is set to \texttt{TRUE}, the model only returns the linear problem (objective function and constraints) of the first stage, as it would be passed to function \texttt{lp} of package \textbf{lpSolve}. Note that, in this case, the solution is not computed.

\end{itemize}

Now, we can run a model for the \texttt{Fortune500} dataset, which was already defined in the object \texttt{dataFortune} of class \texttt{deadata}. For instance, for the input-oriented CCR model, we can use

\begin{verbatim}
R> ccrFortune <- model_basic(dataFortune, orientation = "io", rts = "crs")
\end{verbatim}
while for the BCC model of the same characteristics, we would write 
\begin{verbatim}
R> bccFortune <- model_basic(dataFortune, orientation = "io", rts = "vrs")
\end{verbatim}
although \texttt{orientation = "io"} and \texttt{rts = "crs"} are not necessary because they are the default values. Moreover, note that ``\texttt{datadea =}'' is not necessary in the first field because \texttt{datadea} is always the first parameter in model functions.

\subsection[Extracting the results: The dea class]{Extracting the results: The \texttt{dea} class}

The results delivered by any \texttt{model\_xxx} function is an object of class \texttt{dea} which is basically a list containing the information regarding the data, the call to function \texttt{model\_xxx} and the results obtained for each DMU:
\dirtree{%
.1 dea.
.2 modelname: \textrm{name of the model}..
.2 orientation: \textrm{orientation of the model}..
.2 rts: \textrm{returns to scale of the model}..
.2 DMU: \textrm{results of the model for each evaluated DMU}..
.2 data: \textrm{the object of class \texttt{deadata} to which the model has been applied}..
.2 dmu\_eval: \textrm{evaluated DMUs}..
.2 dmu\_ref: \textrm{evaluation reference set (with respect to which DMUs we have evaluated)}..
.2 maxslack: \textrm{logical parameter indicating if the max-slack solution has been computed}..
.2 weight\_slack\_i: \textrm{weight vector for input slacks in the max-slack solution}..
.2 weight\_slack\_o: \textrm{weight vector for output slacks in the max-slack solution}..
}
Other specific parameters for some models (such as parameters \texttt{L} and \texttt{U} for generalized returns to scale, translation vectors \texttt{vtrans\_i} and \texttt{vtrans\_o} for undesirable variables, or \texttt{orientation\_param}, with the input and output directions in directional models) are also stored because we want the class \texttt{dea} object to contain all the information about the model in question so that the results can be replicated.

The field \texttt{DMU} is itself a list containing, for each one of the evaluated DMUs, all the results obtained by the model. Namely, 
\begin{itemize}
\item \texttt{efficiency}: score (optimal objective value) returned by the model.
\item \texttt{lambda}: optimal $\bm{\lambda}$ vector.
\item \texttt{slack\_input} and \texttt{slack\_output}: optimal slacks.
\item \texttt{target\_input} and \texttt{target\_output}: projection of the evaluated DMU onto the efficient frontier. 
\item \texttt{multiplier\_input} and \texttt{multiplier\_output}: optimal multipliers in multiplier models.
\end{itemize}

In order to easily obtain those aforementioned results, there are several functions designed to extract the required information from the \texttt{dea} class object: \texttt{efficiencies}, \texttt{lambdas}, \texttt{slacks}, \texttt{targets} and \texttt{multipliers}. For example, in the case of the CCR model applied to the \texttt{Fortune500} dataset,

\begin{verbatim}
R> efficiencies(ccrFortune)
\end{verbatim}
\begin{verbatim}
    Mitsubishi         Mitsui         Itochu General Motors       Sumitomo 
       0.66283        1.00000        1.00000        1.00000        1.00000 
      Marubeni     Ford Motor   Toyota Motor          Exxon    Shell Group 
       0.97197        0.73717        0.52456        1.00000        0.84142 
      Walmart         Hitachi      Nippon LI     Nippon T&T           AT&T 
       1.00000        0.38606        1.00000        0.34858        0.27038 
\end{verbatim}
shows the efficiency scores stored in \texttt{ccrFortune}. It is important to remark that function \texttt{efficiencies} returns the scores (i.e., optimal objective values) of the model, that may not always be interpreted as efficiency scores.

Moreover, there are some other functions such as \texttt{references} or \texttt{eff\_dmus}, whose parameter is also a \texttt{dea} object. Function \texttt{references} returns a list with the \emph{reference set} for each inefficient DMU. Note that the \emph{reference set} of a DMU is formed by all the efficient DMUs that appear in the linear combination that conforms its target. On the other hand, function \texttt{eff\_dmus} returns an array with the efficient DMUs evaluated by the corresponding model.

Alternatively, instead of running all the aforementioned functions to extract the results, we can use the function \texttt{summary.dea}, which is a specific method for \texttt{dea} class objects and can be invoked with the generic function \texttt{summary}. For instance, 
\begin{verbatim}
R> res <- summary(ccrFortune, exportExcel = TRUE, returnList = TRUE)
\end{verbatim}
returns all the results stored in \texttt{ccrFortune} as a list of data frames. Otherwise, if \texttt{returnList = FALSE} (by default), then all these data frames are column-wise merged into a single data frame. Anyway, since \texttt{exportExcel = TRUE} (by default), the results in \texttt{res} are also exported to an \textsf{Excel} file named ``\texttt{ResultsDEAYYYYmmdd\_HH.MM.SS.xls}'', where \texttt{YYYYmmdd\_HH.MM.SS} represents the current system date and time. This default name can be changed with the parameter \texttt{filename}.

\begin{figure}[htbp]
  \centering
  \includegraphics[width = \linewidth]{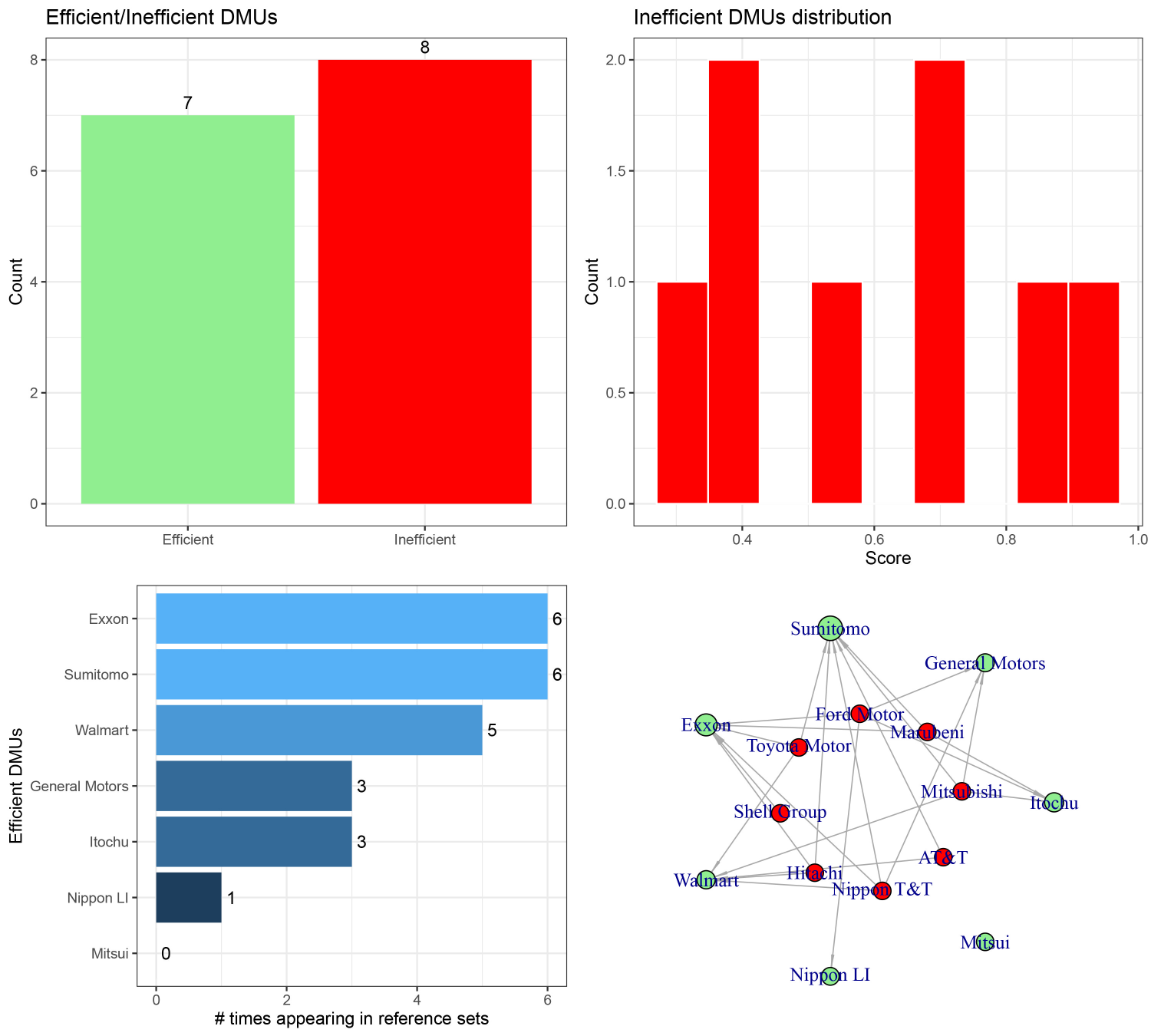}
  \caption{Plots returned by \texttt{plot(ccrFortune)}. In the last plot, efficient DMUs are represented by green circles and inefficient DMUs by red circles. In each inefficient DMU, there are arrows pointing to the DMUs of its corresponding reference set. Moreover, the size of the circle of an efficient DMU depends on the relevance of this DMU in the reference sets.}
  \label{fig:plotccr}
\end{figure}

Finally, we can make use of the function \texttt{plot}, which needs a \texttt{dea} class object in order to make some plots, depending on the model. For example, \texttt{plot(ccrFortune)} returns the plots shown in Figure \ref{fig:plotccr}.

\section{Other models}
\label{sec:models}

In this section, we are going to review the rest of the DEA models implemented in functions of the form \texttt{model\_xxx}. Analogously to \texttt{model\_basic}, most of these functions also use as input parameters \texttt{datadea}, \texttt{orientation}, \texttt{rts}, \texttt{dmu\_eval}, \texttt{dmu\_ref}, \texttt{maxslack}, \texttt{weight\_slack\_i}, \texttt{weight\_slack\_o} and \texttt{returnlp}.

\subsection{Multiplier models}
All basic radial models can be applied in multiplier form \citep{Charnes1962}, solving the dual of the linear problem in the first stage. For example, the input and output-oriented GRS models in multiplier form are given by
\begin{equation}
\label{eq:mult}
\def\arraystretch{1.2}
\begin{array}{ll}
\begin{array}[t]{rl}\textrm{(a)\qquad} \\
\max \limits_{\mathbf{v},\mathbf{u},\xi_L,\xi_U} & \mathbf{u}\mathbf{y}_o+L\xi _L+U\xi _U \\
\text{s.t.} & \mathbf{v} \mathbf{x}_o=1, \\
& -\mathbf{v}X+\mathbf{u}Y+(\xi _L+\xi _U)\mathbf{e} \leq \mathbf{0}, \\
& \mathbf{v}\geq \mathbf{0},\,\, \mathbf{u}\geq \mathbf{0},\,\, \xi_L\geq0,\,\, \xi_U\leq 0,
\end{array}
&
\begin{array}[t]{rl}\textrm{(b)\qquad} \\
\min \limits_{\mathbf{v},\mathbf{u},\xi_L,\xi_U} & \mathbf{v}\mathbf{x}_o+L\xi _L+U\xi _U \\
\text{s.t.} & \mathbf{u} \mathbf{y}_o=1, \\
& \mathbf{v}X-\mathbf{u}Y+(\xi _L+\xi _U)\mathbf{e} \geq \mathbf{0}, \\
& \mathbf{v}\geq \mathbf{0},\,\, \mathbf{u}\geq \mathbf{0},\,\, \xi _L\leq0,\,\, \xi _U\geq 0,
\end{array}
\end{array}
\end{equation}
respectively, where $\mathbf{e}=(1,\ldots,1)$ is a row vector, $\mathbf{v}$ and $\mathbf{u}$ are row vectors interpreted as input and output ``weights'' respectively, and $\xi _L$, $\xi _U$ are the multipliers associated to the returns to scale constraints.
Models with different returns to scale can be deduced from \eqref{eq:mult} by taking $\xi _L=\xi _U=0$ (CRS), $L=U=1$ (VRS), $\xi _L=0$, $U=1$ (NIRS) or $L=1$, $\xi _U=0$ (NDRS).
In any case, $\textrm{DMU}_o$ is \emph{efficient} if and only if the optimal objective of \eqref{eq:mult} (a) or (b) is equal to $1$ and there exists at least one optimal solution with positive optimal weights $\mathbf{v}^*>\mathbf{0}$, $\mathbf{u}^*>\mathbf{0}$. For this reason, the zeroes in the non-negativity conditions of $\mathbf{v}$ and $\mathbf{u}$ are usually replaced by a positive non-Archimedean infinitesimal $\epsilon $.

Multiplier models are applied using \texttt{model\_multiplier}. For example,
\begin{verbatim}
R> multiplierFortune <- model_multiplier(dataFortune, epsilon = 1e-6)
R> multFortune <- multipliers(multiplierFortune)
R> efficiencies(multiplierFortune)
\end{verbatim}
\begin{verbatim}
    Mitsubishi         Mitsui         Itochu General Motors       Sumitomo 
       0.65578        0.96298        1.00000             NA        1.00000 
      Marubeni     Ford Motor   Toyota Motor          Exxon    Shell Group 
       0.95908             NA        0.48254        1.00000        0.82191 
      Walmart         Hitachi      Nippon LI     Nippon T&T           AT&T 
       0.48230        0.29434        1.00000        0.30271        0.19325 
\end{verbatim}
solves the CRS input-oriented multiplier model applied to the \texttt{Fortune500} dataset, stores the multipliers in a list named \texttt{multFortune} and shows the efficiency scores.  Note that \texttt{rts = "crs"} and \texttt{orientation = "io"} are not necessary because they are the default values. Parameter \texttt{epsilon} is the non-Archimedean infinitesimal $\epsilon$ whose default value is $0$. It is important to remark that a too high positive value for \texttt{epsilon} can significantly alter the results and produce infeasibilities. Precisely, in our example, ``General Motors'' and ``Ford Motor'' have \texttt{NA} efficiency scores  because \texttt{epsilon} is too high and the model becomes infeasible. On the other hand, we can also use \texttt{model\_basic} with parameter \texttt{compute\_multiplier = TRUE}, but in this case we can not set parameter \texttt{epsilon}, which is taken as $0$.

\subsection{Free disposal hull models}

Free disposal hull (FDH) models consider the production possibility set
\begin{equation}
\label{eq:pfdh}
P _{FDH}=\left\{ \left( \mathbf{x},\mathbf{y} \right) \in \mathbb{R}_{>0}^{m+s}\ \ | \ \ \mathbf{x}\geq \mathbf{x}_j,\quad \mathbf{y}\leq \mathbf{y}_j,\quad j=1,\ldots,n \right\} ,
\end{equation}
where $\mathbf{x}_j=(x_{1j},\ldots,x_{mj})^\top$ and $\mathbf{y}_j=(y_{1j},\ldots,y_{sj})^\top$ are column vectors. In fact, it is equivalent to consider VRS with $\lambda_j\in \left\{ 0,1\right\} $ binary variables \citep{Thrall1999, Cherchye2000, Deprins2006}. In \textbf{deaR}, these models are applied using \texttt{model\_fdh}, considering the production possibility set \eqref{eq:pfdh} into the models implemented in \texttt{model\_basic}, including directional models (see Section \ref{secdir}). For example, we can replicate some results in \cite{Mamizadeh2013} for the input-oriented case:

\begin{verbatim}
R> dataSupply <- make_deadata(Supply_Chain, inputs = 2:4, outputs = 5:6)
R> fdhSupply <- model_fdh(dataSupply, orientation = "io")
R> efficiencies(fdhSupply)
\end{verbatim}
\begin{verbatim}
    sc1     sc2     sc3     sc4     sc5     sc6     sc7     sc8     sc9 
1.00000 1.00000 0.81550 1.00000 1.00000 1.00000 0.87316 0.73773 1.00000 
   sc10    sc11    sc12    sc13    sc14    sc15    sc16    sc17 
0.84128 0.95342 1.00000 1.00000 1.00000 1.00000 0.98156 1.00000
\end{verbatim}

Note that the efficiency score of DMU \texttt{sc16} is not $1$, contrary to the results shown in  \cite{Mamizadeh2013}.

\subsection{Directional models}
\label{secdir}

The orientation can also be generalized, leading to directional models as described in \cite{Chambers1996,Chambers1998}. The associated linear program for CRS and its second stage are given by
\begin{equation}
\label{eq:dir}
\def\arraystretch{1.2}
\begin{array}{ll}
\begin{array}[t]{rl}
\textrm{(a) }\max \limits_{\beta, \bm{\lambda}} & \beta \\
\text{s.t.} & \beta \mathbf{g}^-+X\bm{\lambda}\leq \mathbf{x}_o, \\
& -\beta \mathbf{g}^++Y\bm{\lambda} \geq \mathbf{y}_o, \\
& \bm{\lambda}\geq \mathbf{0},
\end{array}
&\qquad \qquad
\begin{array}[t]{rl}
\textrm{(b) }\max \limits_{\bm{\lambda}, \mathbf{s}^-,\mathbf{s}^+} & \omega = \mathbf{w}^-\mathbf{s}^-+\mathbf{w}^+\mathbf{s}^+ \\
\text{s.t.} & X\bm{\lambda}+\mathbf{s}^-=\mathbf{x}_o-\beta ^*\mathbf{g}^-, \\
& Y\bm{\lambda} -\mathbf{s}^+=\mathbf{y}_o+\beta ^*\mathbf{g}^+, \\
& \bm{\lambda}\geq \mathbf{0},\,\, \mathbf{s}^-\geq \mathbf{0},\,\, \mathbf{s}^+\geq \mathbf{0},
\end{array}
\end{array}
\end{equation}
respectively, where $\mathbf{g}=(-\mathbf{g}^-,\mathbf{g}^+)\neq \mathbf{0}$ is a preassigned direction (with $\mathbf{g}^-\in \mathbb{R}^m$ and $\mathbf{g}^+\in\mathbb{R}^s$ column vectors), while the weights $\mathbf{w}^-$ and $\mathbf{w}^+$ are positive row vectors. Different returns to scale can be easily considered by adding the corresponding constraints: $\mathbf{e}\bm{\lambda}=1$ (VRS), $0\leq \mathbf{e}\bm{\lambda}\leq 1$ (NIRS), $\mathbf{e}\bm{\lambda}\geq 1$ (NDRS) or $L\leq \mathbf{e}\bm{\lambda}\leq U$ (GRS), with $0\leq L\leq 1$ and $U\geq 1$. Efficient DMUs are those with optimal objectives $\beta ^*=0$ and $\omega ^*=0$. For inefficient DMUs, targets are also given by $(X\bm{\lambda}^*,Y\bm{\lambda}^*)$. If $\mathbf{g}=(-\mathbf{x}_o,\mathbf{0})$ then the model is input-oriented and $\beta ^*=1-\theta ^*$. On the other hand, if $\mathbf{g}=(\mathbf{0},\mathbf{y}_o)$ then the model is output-oriented and $\beta ^*=\eta ^*-1$. Moreover, if $\mathbf{g}=(-\mathbf{x}_o,\mathbf{y}_o)$ then the model is non-oriented and $\beta ^*$ coincides with the \textit{generalized Farrell measure} \citep{Briec1997}.

In function \texttt{model\_basic}, directional models are selected by setting parameter \texttt{orienta\-tion = "dir"}. Then, we can set the input and output directions $\mathbf{g}^-,\mathbf{g}^+$ of each DMU by means of parameters \texttt{dir\_input} and \texttt{dir\_output} respectively which, in general, are matrices of size [number of inputs/outputs]$\times$[number of DMUs in \texttt{dmu\_eval}]. If \texttt{dir\_input} or \texttt{dir\_output} are omitted, then they are assumed to be the input or output matrices (of DMUs in \texttt{dmu\_eval}), respectively. Moreover, if \texttt{dir\_input} or \texttt{dir\_output} are vectors of length [number of inputs/outputs] then the same directions are applied to all evaluated DMUs. Finally, if \texttt{dir\_input} or \texttt{dir\_output} are scalars, then the same constant directions are applied to all inputs/outputs and all evaluated DMUs.

In the following example, considering the \texttt{Fortune500} dataset, we compute the generalized Farrell measures and later, we apply an input-oriented directional model that fully contracts the first input, contracts the second input by half of the usual, and does not contract the third input, all for CRS:
\begin{verbatim}
R> gFmFortune <- model_basic(dataFortune, orientation = "dir")
R> dir_input <- c(1, 0.5, 0) * dataFortune[["input"]]
R> dirFortune <- model_basic(dataFortune, orientation = "dir",
+                            dir_input = dir_input, dir_output = 0)
\end{verbatim}

\paragraph{Range directional models.} Other particular cases of directional models are the range directional models (RDM), in which $g^-_i=x_{io}-\min\left\{ x_{i1},\ldots ,x_{in}\right\} $ for $i=1,\ldots ,m$, and $g^+_r=\max\left\{ y_{r1},\ldots ,y_{rn}\right\} -y_{ro}$ for $r=1,\ldots ,s$,
under VRS \citep{Portela2004}. These models are designed to deal with negative data and they are applied using \texttt{model\_rdm}.
Input and output-oriented versions are also considered by taking $\mathbf{g}^+=\mathbf{0}$ and $\mathbf{g}^-=\mathbf{0}$, respectively. In these cases, parameter \texttt{orientation} must be equal to \texttt{"io"} or \texttt{"oo"}, respectively, instead of the default value \texttt{"no"}. For example,
\begin{verbatim}
R> rdmFortune <- model_rdm(dataFortune, orientation = "io")
R> betascores <- efficiencies(rdmFortune)
\end{verbatim}
returns the optimal $\beta ^*$ scores. Moreover, the inverse range directional models (IRDM) are constructed by substituting the non-zero components of the RDM directions by their inverses. They can be applied by setting parameter \texttt{irdm = TRUE}.

\subsection{Non-radial models}

The non-radial models defined by \cite{Fare1978} allow non-proportional reductions/augmentations in inputs/outputs. The input and output-oriented CRS versions are given  by
\begin{equation}
\label{eq:nr}
\def\arraystretch{1.2}
\begin{array}{ll}
\begin{array}[t]{rl}
\textrm{(a) }\min \limits_{\bm{\theta},\bm{\lambda}} &\bar{\theta}=\frac{1}{m}\mathbf{e}\bm{\theta} \\
\text{s.t.} & \textrm{diag}(\bm{\theta})\mathbf{x}_o-X\bm{\lambda}=\mathbf{0}, \\
& Y\bm{\lambda} \geq \mathbf{y}_o, \\
& \bm{\theta}\leq \mathbf{1},\,\, \bm{\lambda}\geq \mathbf{0},
\end{array}
&\qquad \qquad
\begin{array}[t]{rl}
\textrm{(b) }\max \limits_{\bm{\eta},\bm{\lambda}} &\bar{\eta}=\frac{1}{s}\mathbf{e}\bm{\eta} \\
\text{s.t.} & X\bm{\lambda}\leq \mathbf{x}_o, \\
& \textrm{diag}(\bm{\eta})\mathbf{y}_o-Y\bm{\lambda} =\mathbf{0}, \\
& \bm{\eta}\geq \mathbf{1},\,\, \bm{\lambda}\geq \mathbf{0},
\end{array}
\end{array}
\end{equation}
respectively, where $\mathbf{e}=(1,\ldots,1)$ is a row vector of the adequate dimension, $\bm{\theta}=(\theta_1,\ldots ,\theta _m)^\top $ and $\bm{\eta}=(\eta_1,\ldots ,\eta _s)^\top $ are column vectors, while $\textrm{diag}(\bm{\theta})$ and $\textrm{diag}(\bm{\eta})$ are diagonal matrices. Different returns to scale can be easily considered by adding the corresponding constraints. A second stage for the input and output-oriented CRS versions are needed in order to find the max-slack solution, respectively:
\begin{equation}
\label{eq:nr2}
\def\arraystretch{1.2}
\begin{array}{ll}
\begin{array}[t]{rl}
\textrm{(a) }\max \limits_{\bm{\lambda},\mathbf{s}^+} &\omega^+=\mathbf{w}^+\mathbf{s}^+ \\
\text{s.t.} & X\bm{\lambda}=\textrm{diag}(\bm{\theta}^*)\mathbf{x}_o, \\
& Y\bm{\lambda} -\mathbf{s}^+= \mathbf{y}_o, \\
& \bm{\lambda}\geq \mathbf{0},\,\, \mathbf{s}^+\geq \mathbf{0},
\end{array}
&\qquad \qquad
\begin{array}[t]{rl}
\textrm{(b) }\max \limits_{\bm{\lambda},\mathbf{s}^-} &\omega^-=\mathbf{w}^-\mathbf{s}^- \\
\text{s.t.} & X\bm{\lambda}+\mathbf{s}^-=\mathbf{x}_o, \\
& Y\bm{\lambda} =\textrm{diag}(\bm{\eta}^*)\mathbf{y}_o, \\
& \bm{\lambda}\geq \mathbf{0},\,\, \mathbf{s}^-\geq \mathbf{0},
\end{array}
\end{array}
\end{equation}
where the weights $\mathbf{w}^-$ and $\mathbf{w}^+$ are positive row vectors. $\textrm{DMU}_o$ is efficient if and only if the optimal objectives $\bar{\theta}^*=1$ (or $\bar{\eta}^*=1$) and $\omega ^{+*}=0$ (or $\omega ^{-*}=0$). These non-radial models are applied using \texttt{model\_nonradial}. For the max-slack solution, the weights $\mathbf{w}^-$ or $\mathbf{w}^+$ are introduced by means of parameter \texttt{weight\_slack}, that can be a value, a vector of length [number of inputs/outputs], or a matrix of size [number of inputs/outputs]$\times$[number of DMUs in \texttt{dmu\_eval}]. By default, these weights are set to $1$. For example, we can replicate the results in \cite{Wu2011}:

\begin{verbatim}
R> dataHotels <- make_deadata(Hotels, inputs = 2:5, outputs = 6:8)
R> nonradialHotels <- model_nonradial(dataHotels, orientation = "oo",
+                                     rts = "vrs")
R> head(efficiencies(nonradialHotels)) 
\end{verbatim}
\begin{verbatim}
Warning message:
In make_deadata(Hotels, inputs = 2:5, outputs = 6:8) :
  There are data with very different orders of magnitude. Try to redefine
the units of measure or some linear problems may be ill-posed.
\end{verbatim}
\begin{verbatim}
      Room_revenue F&B_revenue Other_revenue mean_eff
GRA      1.13710     1.07245       4.69421  2.30125
AMB      1.00000     1.00000       1.00000  1.00000
IMP      1.00000     1.00000       5.81116  2.60372
GLP      1.00000     1.00000       1.00000  1.00000
EMP      1.00000     1.00000       1.00000  1.00000
RIV      1.00000     1.00000       1.00000  1.00000
\end{verbatim}

Note that after reading the data, a warning message appears because there are data with very different orders of magnitude. Nevertheless, the model is correctly applied and, in this case, there is no need to rescale the data. 

\paragraph{DEA/preference structure models.} Non-radial models are generalized into DEA/pref\-er\-ence structure models \citep{Zhu1996}, that replace the objectives in \eqref{eq:nr} by a weighted sum and remove the constraints $\bm{\theta}\leq \mathbf{1}$ or $\bm{\eta}\geq \mathbf{1}$. These models are applied using \texttt{model\_deaps}. The weights of the objective, called \emph{preference weights}, are introduced by means of the parameter \texttt{weight\_eff}. Moreover, if the logical parameter \texttt{restricted\_eff} is \texttt{TRUE} (by default) then constraints $\bm{\theta}\leq \mathbf{1}$ or $\bm{\eta}\geq \mathbf{1}$ are not removed. For example, we can apply the input-oriented VRS version to the \texttt{Fortune500} dataset, with objective function $\frac{1}{6}\left( \theta _1+2\theta _2+3\theta _3\right) $:
\begin{verbatim}
R> deapsFortune <- model_deaps(dataFortune, rts = "vrs",
+                              weight_eff = c(1, 2, 3))
\end{verbatim}

\subsection{Additive models}

Additive models \citep{Charnes1985} do not distinguish between orientations and do not need a second stage. The CRS version is given by
\begin{equation}
\label{eq:add}
\def\arraystretch{1.2}
\begin{array}{rl}
\max \limits_{\bm{\lambda},\mathbf{s}^-,\mathbf{s}^+} &\omega=\mathbf{w}^-\mathbf{s}^-+\mathbf{w}^+\mathbf{s}^+ \\
\text{s.t.} & X\bm{\lambda}+\mathbf{s}^-=\mathbf{x}_o, \\
& Y\bm{\lambda} -\mathbf{s}^+=\mathbf{y}_o, \\
& \bm{\lambda}\geq \mathbf{0},\,\, \mathbf{s}^-\geq \mathbf{0},\,\, \mathbf{s}^+\geq \mathbf{0},
\end{array}
\end{equation}
where the weights $\mathbf{w}^-$ and $\mathbf{w}^+$ are positive row vectors. Different returns to scale can be easily considered by adding the corresponding constraints. Hence, $\textrm{DMU}_o$ is efficient if and only if the optimal objective $\omega ^*=0$. Although weights must be positive and orientations are not considered, if $\mathbf{w}^+=\mathbf{0}$ then the model can be interpreted as input-oriented and if $\mathbf{w}^-=\mathbf{0}$ then it can be interpreted as output-oriented. But you have to take into account that if the weight of a slack is zero, then this slack is not taken into account by the objective function and hence, inefficient (weakly efficient) DMUs can get $\omega ^*=0$.

Additive models are applied using \texttt{model\_additive}. The weights are introduced by parameters \texttt{weight\_slack\_i} and \texttt{weight\_slack\_o}. Moreover, parameter \texttt{orientation} can be either \texttt{NULL} (by default), \texttt{"io"} ($\mathbf{w}^+=\mathbf{0}$) or \texttt{"oo"} ($\mathbf{w}^-=\mathbf{0}$).

It is important to note that \eqref{eq:add} is not unit-invariant in general. Nevertheless, there are particular cases that are unit-invariant, like the  measure of inefficiency proportions (MIP) model \citep{Cooper1999}. This model takes the weights $w_i^-=1/x_{io}$ and $w^+_r=1/y_{ro}$ under VRS. For example,
\begin{verbatim}
R> inputs <- dataFortune[["input"]]
R> outputs <- dataFortune[["output"]]
R> mipFortune <- model_additive(dataFortune, rts = "vrs",
+                               weight_slack_i = 1 / inputs,
+                               weight_slack_o = 1 / outputs)
\end{verbatim}
for the \texttt{Fortune500} dataset. Another important particular case is the range adjusted measure (RAM) of inefficiencies model \citep{Cooper1999,Cooper2001}, that can be solved for the \texttt{Fortune500} dataset using the following script:
\begin{verbatim}
R> range_i <- apply(inputs, 1, max) - apply(inputs, 1, min)
R> range_o <- apply(outputs, 1, max) - apply(outputs, 1, min)
R> w_range_i <- 1 / (range_i * (dim(inputs)[1] + dim(outputs)[1]))
R> w_range_o <- 1 / (range_o * (dim(inputs)[1] + dim(outputs)[1]))
R> ramFortune <- model_additive(dataFortune, rts = "vrs",
+                               weight_slack_i = w_range_i,
+                               weight_slack_o = w_range_o) 
\end{verbatim}

Another family of additive models, called \textit{additive-Min}, are developed by \citet{Aparicio2007} in order to find the closest targets to the efficient frontier. The CRS version is given by:
\begin{equation}
\label{eq:addmin}
\def\arraystretch{1.2}
\begin{array}{rl}
\min \limits_{\mathbf{s}^-,\mathbf{s}^+} &\omega _{\textrm{min}}=\mathbf{w}^-\mathbf{s}^-+\mathbf{w}^+\mathbf{s}^+ \\
\text{s.t.} & (\mathbf{x}_o-\mathbf{s}^-,\mathbf{y}_o+\mathbf{s}^+)\textrm{ efficient}, \\
& \mathbf{s}^-\geq \mathbf{0},\,\, \mathbf{s}^+\geq \mathbf{0}.
\end{array}
\end{equation}
Different returns to scale can be considered easily adding the corresponding constraints. However, independently of the returns to scale, these models can produce non-monotonic scores.

Additive-Min models are applied using \texttt{model\_addmin}. Program \eqref{eq:addmin} can be solved by the ``MILP'' method proposed by \citet{Aparicio2007} or the ''maximal friends'' method proposed by \citet{Tone2010}. We can choose the method by means of the parameter \texttt{method}, that can be equal to \texttt{"milp"} or \texttt{"mf"}. We have to take into account that the ``MILP'' method is faster but very problematic numerically. Moreover, for the ``MILP'' method under non-constant returns to scale, a modification proposed by \citet{Zhu2018} is implemented.

In order to apply the ``MILP'' method, we have to compute a set of \textit{extreme efficient} DMUs \citep{Charnes1991}, i.e. DMUs spanning the facets of the efficient frontier that cannot be expressed as a positive linear combination of the other DMUs. We can find a set of extreme efficient DMUs from a \texttt{deadata} object by means of the function \texttt{extreme\_efficient}, and pass this result to \texttt{model\_addmin} through the parameter \texttt{extreff}. On the other hand, if we do not previously compute a set of extreme efficient DMUs, it is computed internally by \texttt{model\_addmin}. For example, we compute the CRS additive-Min scores of the Fortune 500 dataset:
\begin{verbatim}
R> extreffFortune <- extreme_efficient(dataFortune)
R> addminFortune <- model_addmin(dataFortune, extreff = extreffFortune)
R> efficiencies(addminFortune)
\end{verbatim} 
 
On the other hand, for applying the ``maximal friends'' method, we have to previously compute the \textit{maximal friends subsets} \citep{Tone2010}, i.e. the facets of the efficient frontier, by means of function \texttt{maximal\_friends} (see Section \ref{sec:sbm} for more details) and pass the result to \texttt{model\_addmin} through the parameter \texttt{maxfr}.

\subsection{SBM models}
\label{sec:sbm}

Slacks-based measure (SBM) of efficiency models \citep{Tone2001} provide an efficiency score and the CRS version is given by
\begin{equation}
\label{eq:sbmeff}
\def\arraystretch{1.2}
\begin{array}{rl}
\min \limits_{\bm{\lambda},\mathbf{s}^-,\mathbf{s}^+} &\rho =\dfrac{1-\frac{1}{m}\sum _{i=1}^{m}w_i^-s_i^-/x_{io}}{1+\frac{1}{s}\sum _{r=1}^{s}w_r^+s_r^+/y_{ro}} \\
\text{s.t.} & X\bm{\lambda}+\mathbf{s}^-=\mathbf{x}_o, \\
& Y\bm{\lambda} -\mathbf{s}^+=\mathbf{y}_o, \\
& \bm{\lambda}\geq \mathbf{0},\,\, \mathbf{s}^-\geq \mathbf{0},\,\, \mathbf{s}^+\geq \mathbf{0},
\end{array}
\end{equation}
where the weights $w_i^-$, $w_r^+$ are positive with $\sum _{i=1}^mw_i^-=m$ and $\sum _{r=1}^sw_r^+=s$. Note that \eqref{eq:sbmeff} is expressed in a unit-invariant form and, moreover, it can be linearized using the Charnes-Cooper transformation. $\textrm{DMU}_o$ is efficient if and only if  the optimal objective $\rho ^*=1$, i.e., the optimal slacks are all zero.
The input and output-oriented CRS versions are given by:
\begin{equation}
\label{eq:sbmeffor}
\def\arraystretch{1.2}
\begin{array}{ll}
\begin{array}[t]{rl}
\textrm{(a) }\min \limits_{\bm{\lambda},\mathbf{s}^-} &\rho _I=
1-\frac{1}{m}\sum _{i=1}^{m}w_i^-s_i^-/x_{io} 
\\
\text{s.t.} & X\bm{\lambda}+\mathbf{s}^-=\mathbf{x}_o, \\
& Y\bm{\lambda} \geq \mathbf{y}_o, \\
& \bm{\lambda}\geq \mathbf{0},\,\, \mathbf{s}^-\geq \mathbf{0},
\end{array}
&
\begin{array}[t]{rl}
\textrm{(b) }\min \limits_{\bm{\lambda},\mathbf{s}^+} &\rho _O=
1/(1+\frac{1}{s}\sum _{r=1}^{s}w_r^+s_r^+/y_{ro})\\
\text{s.t.} & X\bm{\lambda}\leq \mathbf{x}_o, \\
& Y\bm{\lambda} -\mathbf{s}^+=\mathbf{y}_o, \\
& \bm{\lambda}\geq \mathbf{0},\,\, \mathbf{s}^+\geq \mathbf{0},
\end{array}
\end{array}
\end{equation}
respectively. In general, $\rho ^*_I\geq \rho ^*$ and $\rho ^*_O\geq \rho ^*$. Note that oriented SBM models \eqref{eq:sbmeffor} do not serve to find efficient DMUs by their own because there can be inefficient DMUs with $\rho ^*_I=1$ or $\rho ^*_O=1$. As usual, different returns to scale can be easily considered in \eqref{eq:sbmeff} and \eqref{eq:sbmeffor} by adding the corresponding constraints. SBM models are applied using \texttt{model\_sbmeff}. Parameter \texttt{orientation} can be \texttt{"no"} (non-oriented, by default), \texttt{"io"} (input-oriented) or \texttt{"oo"} (output-oriented). The weights are introduced by means of \texttt{weight\_input} and \texttt{weight\_output}, whose default values are $1$. Moreover, according to \cite{Tone2001}, SBM models in function \texttt{model\_sbmeff} are automatically adapted to deal with zeros in data. We have to note that, for the specific case of zeros in output data, the Case 2 of \citet[p.~507]{Tone2001} is applied, but taking $1/100$ instead of $1/10$.

For example, we can replicate the results in \cite{Tone2001}:

\begin{verbatim}
R> dataTone <- make_deadata(Tone2001, ni = 2, no = 2)
R> sbmTone <- model_sbmeff(dataTone, orientation = "no", rts = "crs")
R> efficiencies(sbmTone)
\end{verbatim}
\begin{verbatim}
  DMU_A   DMU_B   DMU_C   DMU_D   DMU_E 
0.79798 0.56818 1.00000 0.66667 1.00000 
\end{verbatim}

The original SBM efficiency model given by \eqref{eq:sbmeff} evaluates the inefficiency of a DMU referring to the efficient activity of the form $ \left( \mathbf{x}_o-\mathbf{s}^-,\mathbf{y}_o+\mathbf{s}^+\right) $ that produces the lowest score $\rho$. Hence, efficient targets may be far away from $\textrm{DMU}_o$ and they could be inappropriate. To overcome this issue, \cite{Tone2010} among others proposed to search the efficient activity of the form $ \left( \mathbf{x}_o-\mathbf{s}^-,\mathbf{y}_o+\mathbf{s}^+\right) $ that produces the highest score $\rho$, leading to the SBM-Max efficiency model:
\begin{equation}
\label{eq:sbm-max}
\def\arraystretch{1.2}
\begin{array}{rl}
\max\limits_{\mathbf{s}^-,\mathbf{s}^+} & \rho = \dfrac{1-\frac{1}{m}\sum _{i=1}^{m}w_i^-s_i^- /x_{io}}{1+\frac{1}{s}\sum _{r=1}^{s}w_r^+s_r^+ /y_{ro}} \\
\text{s.t.} & \left( \mathbf{x}_o-\mathbf{s}^-,\mathbf{y}_o+\mathbf{s}^+\right) \textrm{ efficient},\\
& \mathbf{s}^-\geq \mathbf{0},\,\, \mathbf{s}^+\geq \mathbf{0}.
\end{array}
\end{equation}
Program \eqref{eq:sbm-max} is solved by means of the ``maximal friends'' (facets of the efficient frontier) technique \citep{Tone2010}. Nevertheless, you may be careful because the SBM-Max efficiency model can produce non-monotonic scores. This model is applied setting the parameter \texttt{kaizen = TRUE} in \texttt{model\_sbmeff}. For example, we can compare the CRS versions of the SBM-Max and the SBM-Min (original) models:

\begin{verbatim}
R> sbmmaxTone <- model_sbmeff(dataTone, kaizen = TRUE, silent = TRUE)
R> efficiencies(sbmmaxTone)
\end{verbatim}
\begin{verbatim}
  DMU_A   DMU_B   DMU_C   DMU_D   DMU_E 
0.86014 0.73427 1.00000 0.66667 1.00000 
\end{verbatim}

Moreover, you can find the maximal friends subsets of a given set of DMUs by means of the function \texttt{maximal\_friends}. The result is a list with all the facets of the efficient frontier and the DMUs that compose them. Moreover, you can pass this result to function \texttt{model\_sbmeff} through the parameter \texttt{maxfr}. For example,
\begin{verbatim}
R> facetsFortune <- maximal_friends(dataFortune, silent = TRUE)
R> sbmmaxFortune <- model_sbmeff(dataFortune, kaizen = TRUE,
+                                maxfr = facetsFortune)
R> efficiencies(sbmmaxFortune)
\end{verbatim}
Parameter \texttt{silent} in functions \texttt{model\_sbmeff} and \texttt{maximal\_friends} allows to hide the progress messages from the computation of the maximal friends. 

\subsection{Cost, revenue and profit models}

The CRS cost, revenue and profit efficiency models \citep{Coelli2005} are given, respectively, by
\begin{equation}
\label{eq:cost}
\def\arraystretch{1.2}
\begin{array}{lll}
\begin{array}[t]{rl}
\textrm{(a) }\min \limits_{\mathbf{x},\bm{\lambda}} &\mathbf{c}\mathbf{x} \\
\text{s.t.} & \mathbf{x}-X\bm{\lambda}\geq \mathbf{0}, \\
& Y\bm{\lambda} \geq \mathbf{y}_o, \\
& \bm{\lambda}\geq \mathbf{0},
\end{array}
&
\begin{array}[t]{rl}
\textrm{(b) }\max \limits_{\mathbf{y},\bm{\lambda}} &\mathbf{p}\mathbf{y} \\
\text{s.t.} & X\bm{\lambda}\leq \mathbf{x}_o, \\
& \mathbf{y}-Y\bm{\lambda} \leq \mathbf{0}, \\
& \bm{\lambda}\geq \mathbf{0},\,\, \mathbf{y}\geq \mathbf{0},
\end{array}
&
\begin{array}[t]{rl}
\textrm{(c) }\max \limits_{\mathbf{x},\mathbf{y},\bm{\lambda}} &\mathbf{p}\mathbf{y}-\mathbf{c}\mathbf{x} \\
\text{s.t.} & \mathbf{x}-X\bm{\lambda}\geq \mathbf{0}, \\
& \mathbf{y}-Y\bm{\lambda} \leq \mathbf{0}, \\
& \mathbf{x}\leq \mathbf{x}_o,\,\, \mathbf{y}\geq \mathbf{y}_o, \\
& \bm{\lambda}\geq \mathbf{0},
\end{array}
\end{array}
\end{equation}
where $\mathbf{c}$ and $\mathbf{p}$ are row vectors with the unit prices of inputs and outputs, respectively. Restricted versions of the cost and revenue efficiency models are given by adding the constraints $\mathbf{x}\leq \mathbf{x}_o$ to \eqref{eq:cost} (a) and $\mathbf{y}\geq \mathbf{y}_o$ to \eqref{eq:cost} (b). The \emph{cost}, \emph{revenue} and \emph{profit efficiency scores} are given, respectively, by
\begin{equation*}
\label{eq:costeff}
\begin{array}{lll}
\textrm{(a) }\dfrac{\mathbf{c}\mathbf{x}^*}{\mathbf{c}\mathbf{x}_o},
&\qquad \qquad
\textrm{(b) }\dfrac{\mathbf{p}\mathbf{y}_o}{\mathbf{p}\mathbf{y}^*},
&\qquad \qquad
\textrm{(c) }\dfrac{\mathbf{p}\mathbf{y}_o-\mathbf{c}\mathbf{x}_o}{\mathbf{p}\mathbf{y}^*-\mathbf{c}\mathbf{x}^*}.
\end{array}
\end{equation*}
$\textrm{DMU}_o$ is considered to be \emph{cost}, \emph{revenue} or \emph{profit efficient} if its respective efficiency score is equal to $1$. Moreover, all efficiency scores are between $0$ and $1$, except for the case $\mathbf{p}\mathbf{y}_o<\mathbf{c}\mathbf{x}_o$, in which the profit efficiency score can be negative or $\geq 1$ (in this case, the higher the score, the greater the inefficiency). Different returns to scale can be easily considered by adding the corresponding constraints.

These models are applied using \texttt{model\_profit}. Unit prices $\mathbf{c}$ and $\mathbf{p}$ are introduced by parameters \texttt{price\_input} and \texttt{price\_output}, respectively. As usual, they can be a value, a vector or a matrix of size [number of inputs/outputs]$\times$[number of DMUs in \texttt{dmu\_eval}]. Restricted versions are considered setting parameter \texttt{restricted\_optimal = TRUE} (by default). For example, the CRS restricted models can be solved for the \texttt{Coelli\_1998} dataset with this script:
\begin{verbatim}
R> dataCoelli <- make_deadata(Coelli_1998, ni = 2, no = 1)
R> price_i <- t(Coelli_1998[, 5:6]) 
R> price_o <- t(Coelli_1998[, 7])
R> costCoelli <- model_profit(dataCoelli, price_input = price_i)
R> revenueCoelli <- model_profit(dataCoelli, price_output = price_o)
R> profitCoelli <- model_profit(dataCoelli, price_input = price_i,
+                               price_output = price_o) 
\end{verbatim}

\section{Special features on variables}
\label{secund}

We can determine with function \texttt{make\_deadata} whether any of the inputs or outputs are either \emph{non-controllable}, \emph{non-discretionary} or \emph{undesirable} variables (see \citet{Cooper2007} and \citet{Zhu2014} for more details about these special features). It can be done with the parameters \texttt{nc\_inputs/nc\_outputs}, \texttt{nd\_inputs/nd\_outputs} and \texttt{ud\_inputs/ud\_outputs}, which are integer numbers denoting the position of non-controllable, non-discretionary and undesirable inputs/outputs, respectively. For example, let us assume that the ``Employees'' input of the \texttt{Fortune500} dataset cannot be controlled by the decision maker, then we can flag this input as non-controllable by

\begin{verbatim}
R> dataFortuneNC <- make_deadata(Fortune500, ni = 3, no = 2, nc_inputs = 3)
\end{verbatim}

Note that ``\texttt{nc\_inputs = 3}'' does not mean ``the third column is non-controllable'', but rather ``the third input (Employees) is non-controllable''.

\subsection{Non-controllable variables}

\emph{Non-controllable} variables can not change their values. The input-oriented CCR model \eqref{eq:ccr} (a) and its second stage \eqref{eq:ccr2} (a) are adapted:
\begin{equation}
\label{eq:ccrnc}
\def\arraystretch{1.2}
\begin{array}{ll}
\begin{array}[t]{rl}
\textrm{(a) }\min \limits_{\theta, \bm{\lambda}} & \theta \\
\text{s.t.} & \theta \mathbf{x}_o^C - X^C\bm{\lambda}\geq \mathbf{0}, \\
& Y^C\bm{\lambda} \geq \mathbf{y}_o^C, \\
& X^{NC}\bm{\lambda} = \mathbf{x}_o^{NC}, \\
& Y^{NC}\bm{\lambda} = \mathbf{y}_o^{NC}, \\
& \bm{\lambda}\geq \mathbf{0},
\end{array}
&\qquad \qquad
\begin{array}[t]{rl}
\textrm{(b) }\max \limits_{\bm{\lambda},\mathbf{s}^{C-},\mathbf{s}^{C+}} &\omega = \mathbf{w}^{C-}\mathbf{s}^{C-}+\mathbf{w}^{C+}\mathbf{s}^{C+} \\
\text{s.t.} & X^C\bm{\lambda}+\mathbf{s}^{C-}=\theta ^*\mathbf{x}_o^C, \\
& Y^C\bm{\lambda} -\mathbf{s}^{C+}=\mathbf{y}_o^C, \\
& X^{NC}\bm{\lambda} = \mathbf{x}_o^{NC}, \\
& Y^{NC}\bm{\lambda} = \mathbf{y}_o^{NC}, \\
& \bm{\lambda}\geq \mathbf{0},\,\, \mathbf{s}^{C-}\geq \mathbf{0},\,\, \mathbf{s}^{C+}\geq \mathbf{0},
\end{array}
\end{array}
\end{equation}
where the superscripts $C$ and $NC$ refers to ``controllable'' and ``non-controllable'' respectively. Analogously, the output-oriented and directional models can be adapted, as well as the other non-radial models. Other returns to scale can be considered by adding the corresponding constraints.

\subsection{Non-discretionary variables}

\emph{Non-discretionary} variables are exogenously fixed and therefore, it is not possible to vary them at the discretion of management \citep{Cooper2007}. The input-oriented CCR model \eqref{eq:ccr} (a) and its second stage \eqref{eq:ccr2} (a) are adapted:
\begin{equation}
\label{eq:ccrnd}
\def\arraystretch{1.2}
\begin{array}{ll}
\begin{array}[t]{rl}
\textrm{(a) }\min \limits_{\theta, \bm{\lambda}} & \theta \\
\text{s.t.} & \theta \mathbf{x}_o^D - X^D\bm{\lambda}\geq \mathbf{0}, \\
& X^{ND}\bm{\lambda} \leq \mathbf{x}_o^{ND}, \\
& Y\bm{\lambda} \geq \mathbf{y}_o, \\
& \bm{\lambda}\geq \mathbf{0},
\end{array}
&\qquad \qquad
\begin{array}[t]{rl}
\textrm{(b) }\max \limits_{\bm{\lambda},\mathbf{s}^{D-},\mathbf{s}^{D+}} &\omega = \mathbf{w}^{D-}\mathbf{s}^{D-}+\mathbf{w}^{D+}\mathbf{s}^{D+} \\
\text{s.t.} & X^D\bm{\lambda}+\mathbf{s}^{D-}=\theta ^*\mathbf{x}_o^D, \\
& X^{ND}\bm{\lambda}\leq \mathbf{x}_o^{ND}, \\
& Y^D\bm{\lambda} -\mathbf{s}^{D+}=\mathbf{y}_o^D, \\
& Y^{ND}\bm{\lambda} \geq \mathbf{y}_o^{ND}, \\
& \bm{\lambda}\geq \mathbf{0},\,\, \mathbf{s}^{D-}\geq \mathbf{0},\,\, \mathbf{s}^{D+}\geq \mathbf{0},
\end{array}
\end{array}
\end{equation}
where the superscripts $D$ and $ND$ refers to ``discretionary'' and ``non-discretionary'' respectively. Analogously, the output-oriented and directional models can be adapted. Other returns to scale can be considered by adding the corresponding constraints. For example, we can replicate the results in \cite{Ruggiero2007}, where the second input is non-discretionary:

\begin{verbatim}
R> dataRuggiero <- make_deadata(Ruggiero2007, ni = 2, no = 1, nd_inputs = 2)
R> ccrRuggiero <- model_basic(dataRuggiero)
R> head(efficiencies(ccrRuggiero))
\end{verbatim}
\begin{verbatim}
   DMU1    DMU2    DMU3    DMU4    DMU5    DMU6 
0.72594 0.88099 0.95681 0.85917 0.97576 0.35795
\end{verbatim}

Another model that can be adapted for non-discretionary variables is the non-radial model by \cite{Fare1978}. For example, the input-oriented CRS non-radial model \eqref{eq:nr} (a) with its second stage \eqref{eq:nr2} (a):
\begin{equation}
\label{eq:nriond}
\def\arraystretch{1.2}
\begin{array}{ll}
\begin{array}[t]{rl}
\textrm{(a) }\min \limits_{\bm{\theta}^D,\bm{\lambda}} &\bar{\theta}^D=\frac{1}{m_1}\mathbf{e}\bm{\theta}^D \\
\text{s.t.} & \textrm{diag}(\bm{\theta}^D)\mathbf{x}_o^D-X^D\bm{\lambda}=\mathbf{0}, \\
& X^{ND}\bm{\lambda} \leq \mathbf{x}_o^{ND}, \\
& Y\bm{\lambda} \geq \mathbf{y}_o, \\
& \bm{\theta}^D\leq \mathbf{1},\,\, \bm{\lambda}\geq \mathbf{0},
\end{array}
&\qquad \qquad
\begin{array}[t]{rl}
\textrm{(b) }\max \limits_{\bm{\lambda},\mathbf{s}^{D+}} &\omega^+=\mathbf{w}^{D+}\mathbf{s}^{D+} \\
\text{s.t.} & X^D\bm{\lambda}=\textrm{diag}(\bm{\theta}^{D*})\mathbf{x}_o^D, \\
& X^{ND}\bm{\lambda} \leq \mathbf{x}_o^{ND}, \\
& Y^D\bm{\lambda} -\mathbf{s}^{D+}=\mathbf{y}_o^D, \\
& Y^{ND}\bm{\lambda} \geq \mathbf{y}_o^{ND}, \\
& \bm{\lambda}\geq \mathbf{0},\,\, \mathbf{s}^{D+}\geq 0.
\end{array}
\end{array}
\end{equation}
Other non-radial models such as additive or SBM are not affected by non-discretionary variables.

\subsection{Undesirable variables}

An output is \emph{undesirable} if producing less quantity of this output leads to more efficiency. By extension, an input is said to be \emph{undesirable} if it behaves contrary to the other usual inputs, i.e., consuming more quantity of this input leads to more efficiency. Note that ``undesirable'' inputs are in fact ``desirable''. A more grammatically correct denomination would be ``good inputs'' and ``bad outputs'', but the term ``undesirable'' prevails in the literature. The production possibility set under CRS is defined by
\begin{equation}
\label{eq:pund}
\def\arraystretch{1.2}
\begin{array}{rcl}
P &=& \left\{ \left( \mathbf{x}^g,\mathbf{x}^b,\mathbf{y}^g,\mathbf{y}^b \right) \in \mathbb{R}_{>0}^{m_1+m_2+s_1+s_2}\ \ | \right. \\
& &\left. \quad \mathbf{x}^g\leq X^g\bm{\lambda},\,\, \mathbf{x}^b\geq X^b\bm{\lambda},\,\, \mathbf{y}^g\leq Y^g\bm{\lambda},\,\, \mathbf{y}^b\geq Y^b\bm{\lambda},\,\, \bm{\lambda}\geq \mathbf{0} \right\} ,
\end{array}
\end{equation}
where the superscripts $g$ and $b$ refers to ``good'' and ``bad'' respectively. Other returns to scale can be considered by adding the corresponding constraints. A $\textrm{DMU}_o$ is efficient in the presence of undesirable inputs/outputs if there is no vector $\left( \mathbf{x}^g,\mathbf{x}^b,\mathbf{y}^g,\mathbf{y}^b \right) \in P$ such that $\mathbf{x}^g_o\leq \mathbf{x}^g$, $\mathbf{x}^b_o\geq \mathbf{x}^b$, $\mathbf{y}^g_o\leq \mathbf{y}^g$, $\mathbf{y}^b_o\geq \mathbf{y}^b$ with at least one strict inequality.

In general, an undesirable output can be treated as an input, and vice versa. Nevertheless, this does not reflect the true production process and there can be interpretation issues in most models. In this case, models must be adapted to undesirable inputs/outputs. For example, a modified version of \eqref{eq:sbmeff} (non-oriented weighted SBM efficiency model under CRS) is
\begin{equation}
\label{eq:sbmeffund}
\def\arraystretch{1.2}
\begin{array}{rl}
\min \limits_{\bm{\lambda},\mathbf{s}^{-},\mathbf{s}^{+}} &\rho =\dfrac{1-\frac{1}{m}\sum _{i=1}^{m}w_i^{-}s_i^{-}/x_{io}}{1+\frac{1}{s}\sum _{r=1}^{s}w_r^{+}s_r^{+}/y_{ro}} \\
\text{s.t.} & X^g\bm{\lambda}-\mathbf{s}^{g-}=\mathbf{x}_o^g, \\
& X^b\bm{\lambda}+\mathbf{s}^{b-}=\mathbf{x}_o^b, \\
& Y^g\bm{\lambda} -\mathbf{s}^{g+}=\mathbf{y}_o^g, \\
& Y^b\bm{\lambda} +\mathbf{s}^{b+}=\mathbf{y}_o^b, \\
& \bm{\lambda}\geq \mathbf{0},\,\, \mathbf{s}^{-}\geq \mathbf{0},\,\, \mathbf{s}^{+}\geq \mathbf{0},
\end{array}
\end{equation}
where $\mathbf{s}^-=(\mathbf{s}^{g-},\mathbf{s}^{b-})$, $\mathbf{s}^+=(\mathbf{s}^{g+},\mathbf{s}^{b+})$, $\mathbf{x}_o=(\mathbf{x}_o^{g},\mathbf{x}_o^{b})$, $\mathbf{y}_o=(\mathbf{y}_o^{g},\mathbf{y}_o^{b})$ \citep{Tone2003,Cooper2007,Tone2021}. Note that some interpretation issues appear in this model because good inputs could generate negative efficiency scores.

Modified versions of the directional CRS model \eqref{eq:dir} (a) and its second stage \eqref{eq:dir} (b) are given in \cite{Fare2004}:
\begin{equation}
\label{eq:dirund}
\def\arraystretch{1.2}
\begin{array}{ll}
\begin{array}[t]{rl}
\textrm{(a) }\max \limits_{\beta, \bm{\lambda}} & \beta \\
\text{s.t.} & -\beta \mathbf{g}^{g-}+X^g\bm{\lambda}= \mathbf{x}_o^g, \\
& \beta \mathbf{g}^{b-}+X^b\bm{\lambda}\leq \mathbf{x}_o^b, \\
& -\beta \mathbf{g}^{g+}+Y^g\bm{\lambda} \geq \mathbf{y}_o^g, \\
& \beta \mathbf{g}^{b+}+Y^b\bm{\lambda} = \mathbf{y}_o^b, \\
& \bm{\lambda}\geq \mathbf{0},
\end{array}
&\qquad \qquad
\begin{array}[t]{rl}
\textrm{(b) }\max \limits_{\bm{\lambda}, \mathbf{s}^{b-},\mathbf{s}^{g+}} & \omega = \mathbf{w}^{b-}\mathbf{s}^{b-}+\mathbf{w}^{g+}\mathbf{s}^{g+} \\
\text{s.t.} & X^b\bm{\lambda}+\mathbf{s}^{b-}=\mathbf{x}_o^b-\beta ^*\mathbf{g}^{b-}, \\
& Y^g\bm{\lambda} -\mathbf{s}^{g+}=\mathbf{y}_o^g+\beta ^*\mathbf{g}^{g+}, \\
& \bm{\lambda}\geq \mathbf{0},\,\, \mathbf{s}^{b-}\geq \mathbf{0},\,\, \mathbf{s}^{g+}\geq \mathbf{0}.
\end{array}
\end{array}
\end{equation}
Other returns to scale can be considered by adding the corresponding constraints.

For non-directional radial models under VRS, undesirable inputs/outputs are treated as proposed by \cite{Seiford2002}. This technique consists of transform each undesirable input/output in this way:
\begin{equation}
^gx_{ij}=-x_{ij}^g+u_i,\qquad ^by_{rj}=-y_{rj}^b+v_r,
\end{equation}
where $\mathbf{u},\mathbf{v}$ are translation vectors that allow $^gx_{ij}$ and $^by_{rj}$ be positive. Usually, the ``max + 1'' translation is applied, i.e., $u_i=\max_{j}\left\{ x_{ij}^g\right\}+1$ and $v_r=\max_{j}\left\{ y_{rj}^b\right\}+1$. The VRS condition is advisable in order to assure translation invariance of the model.
In function \texttt{model\_basic}, parameters \texttt{vtrans\_i} and \texttt{vtrans\_o} correspond to the translation vectors $\mathbf{u}$ and $\mathbf{v}$, respectively. If \texttt{vtrans\_i[i]} is \texttt{NA}, then it applies the ``max + 1'' translation to the i-th undesirable input. If \texttt{vtrans\_i} is a scalar, then it applies the same constant translation to all undesirable inputs. If \texttt{vtrans\_i} is \texttt{NULL}, then it applies the ``max + 1'' translation to all undesirable inputs (analogously for outputs). For example, we can replicate some results in \cite[p.~119]{Hua2007}, where the third output is undesirable and the output translation parameter is set to $1500$:

\begin{verbatim}
R> dataHua <- make_deadata(Hua_Bian_2007, ni = 2, no = 3, ud_outputs = 3)
R> bccHua <- model_basic(dataHua, orientation = "oo", rts = "vrs",
+                        vtrans_o = 1500)
R> head(efficiencies(bccHua))
\end{verbatim}
\begin{verbatim}
   DMU1    DMU2    DMU3    DMU4    DMU5    DMU6 
1.00000 1.00000 1.17726 1.06856 1.00000 1.00000 
\end{verbatim}

Finally, function \texttt{undesirable\_basic} transforms a \texttt{deadata} class object with undesirable inputs/outputs according to \cite{Seiford2002}, making use of parameters \texttt{vtrans\_i} and \texttt{vtrans\_o}. This function also works with \texttt{deadata\_fuzzy} class objects (see Section \ref{sec:readfuzzy}).

\section{Super-efficiency models}
\label{sec:super}

Efficiency models evaluate inefficient DMUs usually providing a score, but they do not discriminate between efficient DMUs. On the other hand, super-efficiency models precisely evaluate efficient DMUs in order to rank them. 

Radial super-efficiency models consist of extracting the evaluated $\mathrm{DMU}_o$ from the evaluation reference set \citep{Andersen1993}. For example, the input-oriented CCR super-efficiency model with its second stage is given by
\begin{equation}
\label{eq:ccrse}
\def\arraystretch{1.2}
\begin{array}{ll}
\begin{array}[t]{rl}
\textrm{(a) }\min \limits_{\theta, \bm{\lambda}} & \theta \\
\text{s.t.} & \theta \mathbf{x}_o - X_{-o}\bm{\lambda}\geq \mathbf{0}, \\
& Y_{-o}\bm{\lambda} \geq \mathbf{y}_o, \\
& \bm{\lambda}\geq \mathbf{0},
\end{array}
&\qquad \qquad
\begin{array}[t]{rl}
\textrm{(b) }\max \limits_{\bm{\lambda},\mathbf{s}^-,\mathbf{s}^+} &\omega = \mathbf{w}^-\mathbf{s}^-+\mathbf{w}^+\mathbf{s}^+ \\
\text{s.t.} & X_{-o}\bm{\lambda}+\mathbf{s}^-=\theta ^*\mathbf{x}_o, \\
& Y_{-o}\bm{\lambda} -\mathbf{s}^+=\mathbf{y}_o, \\
& \bm{\lambda}\geq \mathbf{0},\,\, \mathbf{s}^-\geq \mathbf{0},\,\, \mathbf{s}^+\geq \mathbf{0},
\end{array}
\end{array}
\end{equation}
where $X_{-o},Y_{-o}$ are the input and output data matrices, respectively, defined by $\mathcal{D}-\left\{ \textrm{DMU}_o\right\}$, $\bm{\lambda}=(\lambda_1,\ldots ,\lambda_{o-1},\lambda_{o+1},\ldots ,\lambda_n)^{\top}$, while the weights $\mathbf{w}^-$ and $\mathbf{w}^+$ are positive row vectors. Other returns to scale can be considered, but infeasibility problems may appear. Radial super-efficiency models can be applied using \texttt{model\_supereff}.

With respect to non-radial models, the SBM super-efficiency (SSBM) models consist of projecting the evaluated $\mathrm{DMU}_o$ onto the part of the production possibility set defined by $\mathcal{D}-\left\{ \textrm{DMU}_o\right\}$ that consumes more inputs and produces less outputs than $\mathrm{DMU}_o$ \citep{Tone2002}. For example, the non-oriented CRS version is given by
\begin{equation}
\label{eq:sbmseff}
\def\arraystretch{1.2}
\begin{array}{rl}
\min \limits_{\bm{\lambda},\mathbf{t}^-,\mathbf{t}^+} &\delta =\dfrac{1+\frac{1}{m}\sum _{i=1}^{m}w_i^-t_i^-/x_{io}}{1-\frac{1}{s}\sum _{r=1}^{s}w_r^+t_r^+/y_{ro}} \\
\text{s.t.} & X_{-o}\bm{\lambda}-\mathbf{t}^-\leq \mathbf{x}_o, \\
& Y_{-o}\bm{\lambda} +\mathbf{t}^+\geq \mathbf{y}_o, \\
& \bm{\lambda}\geq \mathbf{0},\,\, \mathbf{t}^-\geq \mathbf{0},\,\, \mathbf{t}^+\geq \mathbf{0},
\end{array}
\end{equation}
where $\mathbf{w}^-=\left( w_1^-,\ldots ,w_m^-\right) $ and $\mathbf{w}^+=\left( w_1^+,\ldots ,w_s^+\right) $ are weights, and $\mathbf{t}^-$, $\mathbf{t}^+$ are called \emph{super-slacks}. Other returns to scale and orientations can be considered \citep{Tone2002}. SSBM models are applied using the function \texttt{model\_sbmsupereff}, in which parameters \texttt{weight\_slack\_i} and \texttt{weight\_slack\_o} stands for $\mathbf{w}^-$ and $\mathbf{w}^+$, respectively. As usual, they can be a value ($1$ by default), a vector or a matrix of size [number of inputs/outputs]$\times$[number of DMUs in \texttt{dmu\_eval}]. For example, we can replicate the results in \cite{Tone2002}, where the CCR super-efficiency and the input-oriented SSBM models are compared:

\begin{verbatim}
R> dataPower <- make_deadata(Power_plants, ni = 4, no = 2)
R> sccrPower <- model_supereff(dataPower, orientation = "io")
R> efficiencies(sccrPower) 
\end{verbatim}
\begin{verbatim}
     D1      D2      D3      D4      D5      D6 
1.02825 2.41667 1.31250 1.62500 2.40257 1.06279 
\end{verbatim}

\begin{verbatim}
R> ssbmPower <- model_sbmsupereff(dataPower, orientation = "io")
R> efficiencies(ssbmPower)
\end{verbatim}
\begin{verbatim}
     D1      D2      D3      D4      D5      D6 
1.01162 1.70833 1.07812 1.15625 1.79881 1.01981
\end{verbatim}

\cite{Du2010} adapted SSBM models to additive models, changing the objective $\delta $ in \eqref{eq:sbmseff} by $\mathbf{w}^-\mathbf{t}^-+\mathbf{w}^+\mathbf{t}^+$. These additive super-efficiency models are considered to be input-oriented if $\mathbf{w}^+=\mathbf{0}$, and output-oriented if $\mathbf{w}^-=\mathbf{0}$; if both weights are non-zero, they are non-oriented. Additive super-efficiency models can be applied using \texttt{model\_addsupereff}, in which, as in SSBM models, parameters \texttt{weight\_slack\_i} and \texttt{weight\_slack\_o} stands for $\mathbf{w}^-$ and $\mathbf{w}^+$, respectively. By default, $w_i^-=1/x_{io}$ and $w_r^+=1/y_{ro}$, making the model unit invariant. Moreover, for comparison purposes, \texttt{model\_addsupereff} also returns the corresponding score $\delta $ of the optimal solution. In fact, function \texttt{efficiencies} returns these $\delta $ scores. For example, we can verify that, taking the default weights, the input-oriented SSBM and additive super-efficiency models are equivalent (note that it does not hold for other orientations in general): 

\begin{verbatim}
R> saddPower <- model_addsupereff(dataPower, orientation = "io")
R> efficiencies(saddPower)
\end{verbatim}
\begin{verbatim}
     D1      D2      D3      D4      D5      D6 
1.01162 1.70833 1.07812 1.15625 1.79881 1.01981 
\end{verbatim}

\section{Cross-efficiency models}
\label{sec:cross}

Cross-efficiency models evaluate the efficiency score of a DMU using the optimal weights of the other DMUs \citep{Doyle1994}. For example, the input and output-oriented GRS \emph{cross-efficiency} of $\textrm{DMU}_k$ based on the weights of $\textrm{DMU}_o$ are given by
\begin{equation}
\label{eq:crosseff}
\begin{array}{ll}
\textrm{(a) }E_{ok}=\dfrac{\mathbf{u}_o^*\mathbf{y}_k+L\xi _{Lo}^*+U\xi _{Uo}^*}{\mathbf{v}_o^*\mathbf{x}_k},
&\qquad \qquad
\textrm{(b) }E_{ok}=\dfrac{\mathbf{v}_o^*\mathbf{x}_k+L\xi _{Lo}^*+U\xi _{Uo}^*}{\mathbf{u}_o^*\mathbf{y}_k},
\end{array}
\end{equation}
where $\mathbf{v}_o^*,\mathbf{u}_o^*,\xi _{Lo}^*,\xi _{Uo}^*$ are optimal according to programs \eqref{eq:mult} (a) or (b), respectively. Expressions for different returns to scale can be deduced from \eqref{eq:crosseff} by taking $\xi _{Lo}^*=\xi _{Uo}^*=0$ (CRS), $L=U=1$ (VRS), $\xi _{Lo}^*=0$, $U=1$ (NIRS) or $L=1$, $\xi _{Uo}^*=0$ (NDRS). Then, the \emph{cross-efficiency score} of $\textrm{DMU}_k$ is given by the average of the $k$-th column of matrix $E$:
\begin{equation}
\label{eq:crosseffe}
e_k=\frac{1}{n}\sum _{o=1}^n E_{ok}.
\end{equation}
Alternatively, $e_k$ can be computed excluding $E_{kk}$, i.e., without self-appraisal. Other interesting scores are the averages of the rows of $E$:
\begin{equation}
\label{eq:crosseffA}
A_o=\frac{1}{n}\sum _{k=1}^n E_{ok},
\end{equation}
that can be also computed with or without self-appraisal. Finally, the \emph{Maverick index} of $\textrm{DMU}_k$ is given by
\begin{equation}
\label{eq:mav}
M_k=\frac{E_{kk}-e_k}{e_k}.
\end{equation}

Since optimal weights may not be unique, values of cross-efficiencies may vary depending on which optimal weights are chosen. In order to avoid this arbitrariness, two additional linear methods (II and III) with two different formulations (aggressive and benevolent) are implemented \citep{Doyle1994}.
For example, the aggressive input-oriented GRS method II applied to $\textrm{DMU}_o$ is given by:
\begin{equation}
\label{eq:m2}
\def\arraystretch{1.2}
\begin{array}{rl}
\min \limits_{\mathbf{v},\mathbf{u},\xi_L,\xi_U} & -\mathbf{v}\sum _{k\neq o}\mathbf{x}_k+\mathbf{u}\sum _{k\neq o}\mathbf{y}_k+(n-1)(L\xi _L+U\xi _U) \\
\text{s.t.} & \mathbf{v} \mathbf{x}_o=1, \\
& -\mathbf{v}X+\mathbf{u}Y+(\xi _L+\xi _U)\mathbf{e} \leq \mathbf{0}, \\
& \mathbf{u}\mathbf{y}_o+L\xi _L+U\xi _U=E_{oo}, \\
& \mathbf{v}\geq \mathbf{0},\,\, \mathbf{u}\geq \mathbf{0},\,\, \xi_L\geq0,\,\, \xi_U\leq 0,
\end{array}
\end{equation}
and the corresponding method III:
\begin{equation}
\label{eq:m3}
\def\arraystretch{1.2}
\begin{array}{rl}
\min \limits_{\mathbf{v},\mathbf{u},\xi_L,\xi_U} & \mathbf{u}\sum _{k\neq o}\mathbf{y}_k+(n-1)(L\xi _L+U\xi _U) \\
\text{s.t.} & \mathbf{v}\sum _{k\neq o}\mathbf{x}_k=1, \\
& -\mathbf{v}X+\mathbf{u}Y+(\xi _L+\xi _U)\mathbf{e} \leq \mathbf{0}, \\
& -E_{oo}\mathbf{v}\mathbf{x}_o+\mathbf{u}\mathbf{y}_o+L\xi _L+U\xi _U=0, \\
& \mathbf{v}\geq \mathbf{0},\,\, \mathbf{u}\geq \mathbf{0},\,\, \xi_L\geq0,\,\, \xi_U\leq 0.
\end{array}
\end{equation}
The benevolent versions of \eqref{eq:m2} and \eqref{eq:m3} are given by maximizing instead of minimizing. Moreover, it is important to remark that the aggressive formulations of methods II and III under non constant returns to scale can lead to unbounded programs. In this case, \textbf{deaR} adds bound constraints automatically.

Finally, the correction proposed by \cite{Lim2015b} can be applied in the input-oriented VRS model in order to fix negative cross-efficiency scores. This correction has been implemented analogously in the input-oriented NIRS and GRS models which can also give negative cross-efficiencies. For example, the corrected input-oriented GRS cross-efficiency of $\textrm{DMU}_k$ based on the weights of $\textrm{DMU}_o$ is given by
\begin{equation}
E_{ok}=\frac{\mathbf{u}_o^*\mathbf{y}_k}{\mathbf{v}_o^*\mathbf{x}_k-L\xi _{Lo}^*-U\xi _{Uo}^*}.
\end{equation}
Method II is not affected by this correction, but method III can be adapted. In this way, program \eqref{eq:m3} becomes
\begin{equation}
\label{eq:m3c}
\def\arraystretch{1.2}
\begin{array}{rl}
\min \limits_{\mathbf{v},\mathbf{u},\xi_L,\xi_U} & \mathbf{u}\sum _{k\neq o}\mathbf{y}_k \\
\text{s.t.} & \mathbf{v}\sum _{k\neq o}\mathbf{x}_k-(n-1)(L\xi _L+U\xi _U)=1, \\
& -\mathbf{v}X+\mathbf{u}Y+(\xi _L+\xi _U)\mathbf{e} \leq \mathbf{0}, \\
& -E_{oo}\mathbf{v}\mathbf{x}_o+\mathbf{u}\mathbf{y}_o+L\xi _L+U\xi _U=0, \\
& \mathbf{v}\geq \mathbf{0},\,\, \mathbf{u}\geq \mathbf{0},\,\, \xi_L\geq0,\,\, \xi_U\leq 0.
\end{array}
\end{equation}

Cross-efficiency models can be applied using \texttt{cross\_efficiency}. Apart from the usual \texttt{datadea}, \texttt{orientation}, \texttt{rts}, \texttt{dmu\_eval} and \texttt{dmu\_ref}, there are some other interesting parameters:
\begin{itemize}
\item \texttt{epsilon}: multipliers must be $\geq \epsilon$ (default value $0$).

\item \texttt{selfapp}: if this logical variable is set to \texttt{TRUE} (by default), self-appraisal is included in the computations of $e_k$ \eqref{eq:crosseffe} and $A_o$ \eqref{eq:crosseffA}.

\item \texttt{correction}: if this logical variable is set to \texttt{TRUE} (default value \texttt{FALSE}), the correction proposed by \cite{Lim2015b} is applied in the input-oriented VRS, NIRS and GRS models.

\item \texttt{M2} and \texttt{M3}: if these logical variables are set to \texttt{TRUE} (by default), methods II and III are computed.
\end{itemize}

The output of \texttt{cross\_efficiency} is a list with fields \texttt{orientation}, \texttt{rts}, \texttt{L}, \texttt{U}, \texttt{selfapp}, \texttt{correction}, \texttt{Arbitrary}, \texttt{M2\_agg}, \texttt{M2\_ben}, \texttt{M3\_agg}, \texttt{M3\_ben}, \texttt{data}, \texttt{dmu\_eval}, \texttt{dmu\_ref}, \texttt{epsi\-lon} and \texttt{modelname}. The results of arbitrary, method II (aggressive and benevolent) and method III (aggressive and benevolent) models are stored in \texttt{Arbitrary}, \texttt{M2\_agg}, \texttt{M2\_ben}, \texttt{M3\_agg} and \texttt{M3\_ben}, respectively. These fields have subfields \texttt{multiplier\_input}, \texttt{multiplier\-\_output}, \texttt{multiplier\_rts} (for non constant returns to scale), \texttt{cross\_eff}, \texttt{e}, \texttt{A} and \texttt{maverick}. Moreover, \texttt{efficiency} is stored in the field \texttt{Arbitrary}.

We can replicate the results in \cite{Golany1989}:
\begin{verbatim}
R> dataGolany <- make_deadata(Golany_Roll_1989, inputs = 2:4, outputs = 5:6)
R> crossGolany <- cross_efficiency(dataGolany)
\end{verbatim}
For example, we can show the cross-efficiency scores \eqref{eq:crosseffe} for the benevolent formulation of method II: 

\begin{verbatim}
R> crossGolany$M2_ben$e
\end{verbatim}
\begin{verbatim}
    DMU_1     DMU_2     DMU_3     DMU_4     DMU_5     DMU_6     DMU_7 
0.5856330 0.7494068 0.5686789 0.8233164 0.4818662 0.5902805 0.6236033 
    DMU_8     DMU_9    DMU_10    DMU_11    DMU_12    DMU_13 
0.5179766 0.3942743 0.7588777 0.9170085 0.9853480 0.9902515 
\end{verbatim}

\begin{figure}[htbp]
  \centering
  \includegraphics[width = \linewidth]{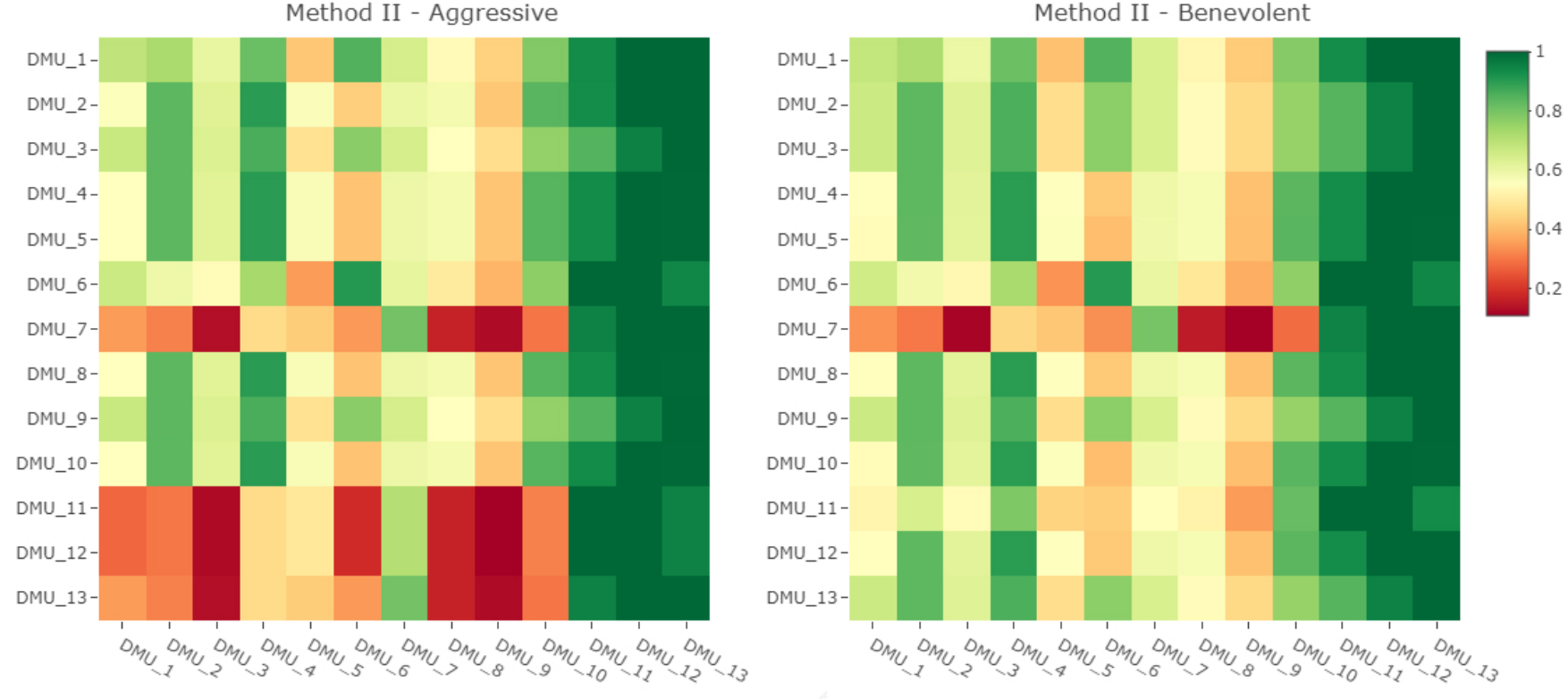}
  \caption{Plots of the cross-efficiency matrices $E$ from aggressive and benevolent formulations of method II, returned by \texttt{plot(crossGolany)}.}
  \label{fig:cross}
\end{figure}

Moreover, we can use function \texttt{plot} in order to visualize cross-efficiency matrices $E$ from different methods and formulations, as shown in Figure \ref{fig:cross}.

Finally, we can replicate the results in \cite{Lim2015b} and compare cross-efficiency scores for different methods:

\begin{verbatim}
R> dataLim <- make_deadata(Lim_Zhu_2015, ni = 1, no = 5)
R> crossLim <- cross_efficiency(dataLim, rts = "vrs", correction = TRUE)
R> head(crossLim$Arbitrary$e)
\end{verbatim}
\begin{verbatim}
Project_1 Project_2 Project_3 Project_4 Project_5 Project_6 
0.7073056 0.6138268 0.1847451 0.4605659 0.4957667 0.5759273 
\end{verbatim}

\begin{verbatim}
R> head(crossLim$M2_agg$e)
\end{verbatim}
\begin{verbatim}
Project_1 Project_2 Project_3 Project_4 Project_5 Project_6 
0.7247253 0.6388864 0.1825081 0.4472565 0.5004328 0.5738613 
\end{verbatim}

\begin{verbatim}
R> head(crossLim$M2_ben$e)
\end{verbatim}
\begin{verbatim}
Project_1 Project_2 Project_3 Project_4 Project_5 Project_6 
0.7484397 0.6486081 0.1977408 0.4975789 0.5305743 0.6249486 
\end{verbatim}

\begin{verbatim}
R> head(crossLim$M3_agg$e)
\end{verbatim}
\begin{verbatim}
Project_1 Project_2 Project_3 Project_4 Project_5 Project_6 
0.6829132 0.5973632 0.1811912 0.4359132 0.4768348 0.5500825 
\end{verbatim}

\begin{verbatim}
R> head(crossLim$M3_ben$e)
\end{verbatim}
\begin{verbatim}
Project_1 Project_2 Project_3 Project_4 Project_5 Project_6 
0.7524549 0.6500783 0.2243236 0.5006576 0.5338293 0.6024692 
\end{verbatim}

Note that bound constraints are automatically added for unbounded methods under non constant returns to scale.

\section{Fuzzy models}
\label{sec:fuzzy}

In all the DEA models seen so far, it is assumed that the data about the production process (i.e., inputs and outputs) are perfectly known, and thus, they can be considered deterministic. However, in practice, it is quite common that some degree of uncertainty is present in the data, and therefore methods that can deal with such vagueness must be defined. 

There are mainly two approaches for dealing with non-deterministic data in DEA: stochastic and fuzzy models. The former uses statistical distributions to obtain a statistical characterization of the efficient frontier (see \cite{Olesen2016} and references therein for a review on stochastic DEA), while the latter uses fuzzy theory and membership functions to deal with the ambiguity in the data (see \cite{Hatami2011,Emrouznejad2014} for reviews on fuzzy DEA). 

Currently, in the \textbf{deaR} package, three popular fuzzy models are implemented, namely: Kao-Liu,  Guo-Tanaka and possibilistic models. Next, we shall describe those models and their implementation in \textbf{deaR}, after giving a brief introduction to fuzzy numbers. 

\subsection{A primer on fuzzy numbers}

A \emph{fuzzy set} $A$ is defined by a function $\mu _A:\mathbb{R}\rightarrow \left[ 0,1\right] $, called \emph{membership function}. 
For any $x\in \mathbb{R}$, the value $\mu _A(x)$ can be interpreted as the grade of membership of $x$ to $A$. Alternatively, a fuzzy set is completely determined by the so-called $\alpha$-\emph{cuts} (or $h$-\emph{levels}) which are defined by $A^{\alpha}=\left\{ x\in \mathbb{R}\,\,|\,\,\mu_A(x) \geq \alpha \right\} $ for $\alpha \in \left] 0,1\right]$, and $A^0=\overline{\left\{ x\in \mathbb{R}\,\,|\,\,\mu_A(x)>0\right\}}$, where the overline denotes clausure.

\begin{figure}[htbp]
  \centering
  \includegraphics[width = .8\linewidth]{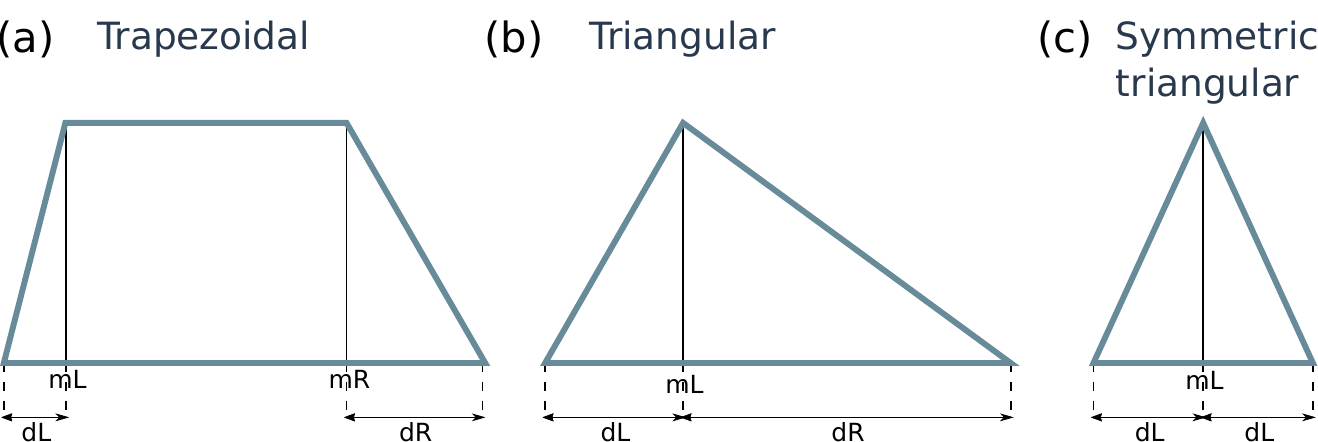}
  \caption{Types of fuzzy numbers considered in package \textbf{deaR} and their respective definition parameters.}
  \label{fig:fuzzy}
\end{figure}

\emph{Fuzzy numbers} are a particular case of fuzzy sets verifying:
\begin{itemize}
\item $A^{\alpha}$ are convex (i.e., intervals) for $\alpha \in \left[ 0,1\right] $.
\item $A^1\neq \emptyset$ (normalized).
\end{itemize}
In \textbf{deaR}, we only consider a special type of fuzzy numbers called \emph{trapezoidal}, whose membership functions have a trapezoidal shape as shown in Figure~\ref{fig:fuzzy} (a). Thus, a trapezoidal fuzzy number is defined by the parameters $(\texttt{mL}, \texttt{mR}, \texttt{dL}, \texttt{dR})$. If $\texttt{mL}=\texttt{mR}$, the fuzzy number is called \emph{triangular}; moreover, if $\texttt{dL}=\texttt{dR}$, it is called \emph{symmetric}. Note that a \emph{crisp number} is a degenerated case of symmetric triangular fuzzy number for which $\texttt{dL}=0$. 

\subsection[Introducing fuzzy data: The deadata fuzzy class]{Introducing data: The \texttt{deadata\_fuzzy} class}
\label{sec:readfuzzy}

Let us assume that our DEA dataset contains some inputs/outputs which are trapezoidal fuzzy numbers. Then, for each one of those inputs/outputs, four columns must be defined, and they can be read with the \texttt{make\_deadata\_fuzzy} function.
Parameters  $(\texttt{mL}, \texttt{mR}, \texttt{dL}, \texttt{dR})$ corresponding to input variables are introduced by \texttt{inputs.mL}, \texttt{inputs.mR}, \texttt{inputs.dL} and \texttt{inputs.dR}, respectively. These parameters are numeric vectors of length $m$ (number of inputs), specifying in which column of the dataset the corresponding quantity is located. If any parameter is not defined for a given input, the corresponding position of the vector must be filled with \texttt{NA}. Analogously for outputs.

For example, \texttt{Kao\_Liu\_2003} dataset \citep{Kao2003} contains one crisp input and five outputs, two of them (3rd and 5th) being triangular fuzzy numbers. This dataset should be read as follows:
\begin{verbatim}
R> dataKao <- make_deadata_fuzzy(Kao_Liu_2003, inputs.mL = 2,
+                                outputs.mL = 3:7, 
+                                outputs.dL = c(NA, NA, 8, NA, 10),
+                                outputs.dR = c(NA, NA, 9, NA, 11))
\end{verbatim}
Note that, in this example, only \texttt{inputs.mL} is needed because the input is crisp. Then, \texttt{inputs.mR} is taken equal to \texttt{inputs.mL}, and \texttt{inputs.dL} $=$ \texttt{inputs.dR} $=0$ automatically. On the other hand, \texttt{outputs.mR} is also not necessary because fuzzy outputs are triangular. The 1st, 2nd and 4th outputs are crisp, hence the 1st, 2nd and 4th entries of vectors \texttt{outputs.dL} and \texttt{outputs.dR} are set to \texttt{NA}. The 3rd and 5th entries of these vectors contain the column positions of the corresponding parameters (e.g., the 8th column of \texttt{Kao\_Liu\_2003} dataset contains the \texttt{dL} parameter of the 3rd output).

For the sake of completeness we are going to read, in the next example, a mixed fuzzy dataset with different types of the aforementioned fuzzy numbers as variables.
The dataset \texttt{FuzzyExample} is included in \textbf{deaR} and contains 5 DMUs with 3 inputs and 3 outputs, whose types are given in Table~\ref{tab:fuzzyextype}. Moreover, in this example, inputs and outputs corresponding to DMUs A and D are crisp numbers, but we need to write them as fuzzy numbers, i.e., $\texttt{mR}=\texttt{mL}$, and $\texttt{dL}=\texttt{dR}=0$ (see Table~\ref{tab:fuzzyexample}). The data reading is as follows:

\begin{table}
\small
\centering
\label{tab:fuzzyextype}
\begin{tabular}{lll}
\toprule
Variable&Type&Definition parameters\\\midrule
Input 1&Crisp&\texttt{mL}.\\
Input 2&Trapezoidal (non-symmetric)&\texttt{mL}, \texttt{mR}, \texttt{dL}, \texttt{dR}\\
Input 3& Symmetric triangular&\texttt{mL}, \texttt{dL}\\
Output 1&Crisp&\texttt{mL}\\
Output 2&Trapezoidal symmetric&\texttt{mL}, \texttt{mR}, \texttt{dL}\\
Output 3&Triangular (non-symmetric)&\texttt{mL}, \texttt{dL}, \texttt{dR}\\
\bottomrule
\end{tabular}
\caption{Input and output types in \texttt{FuzzyExample} dataset with their respective definition parameters.}
\end{table}
 
\begin{verbatim}
R> dataFuzzy <- make_deadata_fuzzy(FuzzyExample, inputs.mL = c(2, 3, 7),
+                                  inputs.mR = c(NA, 4, NA),
+                                  inputs.dL = c(NA, 5, 8),
+                                  inputs.dR = c(NA, 6, NA),
+                                  outputs.mL = c(9, 10 , 13),
+                                  outputs.mR = c(NA, 11, NA),
+                                  outputs.dL = c(NA, 12, 14),
+                                  outputs.dR = c(NA, NA, 15))
\end{verbatim}

\begin{table}[ht]
\small
\centering
\label{tab:fuzzyexample}
\begin{tabular}{crrrrrr}
  \toprule
  DMU & \texttt{Input1.mL} & \texttt{Input2.mL} & \texttt{Input2.mR} & \texttt{Input2.dL} & \texttt{Input2.dR} &\ldots \\ 
  \midrule
 A & 23.00 & 12.00 & 12.00 & 0.00 & 0.00 &\ldots\\ 
 B & 28.00 & 18.00 & 21.00 & 2.00 & 1.50 &\ldots\\ 
 C & 42.00 & 14.00 & 18.00 & 3.50 & 0.80 &\ldots\\ 
 D & 32.00 & 21.00 & 21.00 & 0.00 & 0.00 &\ldots\\ 
 E & 31.00 & 24.00 & 26.00 & 1.50 & 1.20 &\ldots\\ 
   \bottomrule
\end{tabular}
\caption{Excerpt of the \texttt{FuzzyExample} dataset.}
\end{table}

Finally, \texttt{make\_deadata\_fuzzy} returns a \texttt{deadata\_fuzzy} class object, whose structure is the same as a \texttt{deadata} object but including subfields \texttt{mL}, \texttt{mR}, \texttt{dL}, \texttt{dR} in the \texttt{input} and \texttt{output} fields. The \texttt{deadata\_fuzzy} objects are passed to \texttt{modelfuzzy\_xxx} functions in order to perform a DEA fuzzy analysis, as we are going to see in the next sections.

\subsection{Kao-Liu models}

Kao-Liu models \citep{Kao2000a,Kao2000b,Kao2003} consist on applying an existing model in the worst and best scenarios for the evaluated DMUs at each $\alpha$-cut. The worst scenario for a DMU occurs when it consumes the largest possible input amounts and produces the smallest possible output amounts, while, on the contrary, the rest of the DMUs consume the smallest possible input amounts and produce the largest possible output amounts. On the other hand, the best scenario for a DMU occurs when it consumes the smallest possible input amounts and produces the largest possible output amounts, while the rest of the DMUs consume the largest possible input amounts and produce the smallest possible output amounts.

Hence, Kao-Liu models are in fact ``metamodels'' because another usual (crisp) model is needed. In \textbf{deaR}, Kao-Liu models are applied using \texttt{modelfuzzy\_kaoliu} and the underlying model is selected by means of the parameter \texttt{kaoliu\_modelname}, whose possible values are ``basic'', ``additive'', ``addsupereff'', ``deaps'', ``fdh'', ``multiplier'', ``nonradial'', ``profit'', ``rdm'', ``sbmeff'', ``sbmsupereff'' and ``supereff''. Specific parameters for these models, such as orientation or returns to scale, can be also introduced.

\begin{figure}[htbp]
  \centering
  \includegraphics[width = .6\linewidth]{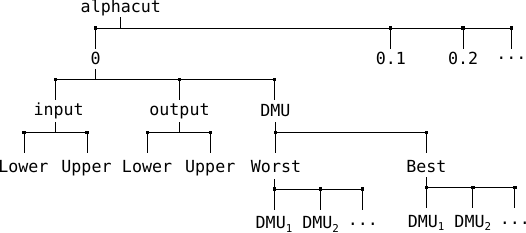}
  \caption{Structure of the field \texttt{alphacut}. Inside of each field \texttt{DMU$_1$}, \texttt{DMU$_2$}, $\ldots$ are stored efficiency scores, lambdas, slacks, targets, multipliers and other results of the submodel.}
  \label{fig:alphatree}
\end{figure}

Finally, $\alpha$-cuts are selected by parameter \texttt{alpha}, that is a numeric vector with the $\alpha$-cuts in $\left[ 0,1\right] $. Alternatively, if \texttt{alpha} $>1$, it determines the number of $\alpha$-cuts, equispatially distributed in $\left[ 0,1\right] $. For example,
\begin{verbatim}
R> kaoliubccKao <- modelfuzzy_kaoliu(dataKao, kaoliu_modelname = "basic",
+                                    alpha = seq(0, 1, by = 0.1),
+                                    orientation = "io", rts = "vrs")  
\end{verbatim}
applies Kao-Liu using the input-oriented BCC model. Note that \texttt{alpha = 11} would produce the same $\alpha$-cuts.
The function \texttt{modelfuzzy\_kaoliu} returns an object of class \texttt{dea\_fuzzy} containing all the information and parameters. The specific results of the applied submodel are stored in the field \texttt{alphacut}. Inside \texttt{alphacut} there are fields for each $\alpha$-cut and, inside these fields, the corresponding input/output data along with the results of each DMU are stored in the field \texttt{DMU}, as shown in Figure~\ref{fig:alphatree}.

One of the difficulties about fuzzy efficiency models is the representation of the efficiency scores. This is particularly cumbersome in Kao-Liu models, since the scores themselves can be non-trapezoidal fuzzy numbers. In the package \textbf{deaR}, some plot methods for objects of class \texttt{dea\_fuzzy} are also implemented in function \texttt{plot}. For example, we can represent the results obtained from the Kao-Liu BCC model applied to the \texttt{Leon\_2003} dataset:

\begin{verbatim}
R> dataLeon <- make_deadata_fuzzy(Leon2003, inputs.mL = 2, inputs.dL = 3, 
+                                 outputs.mL = 4, outputs.dL = 5)
R> kaoliubccLeon <- modelfuzzy_kaoliu(dataLeon, kaoliu_modelname = "basic", 
+                                     alpha = 5,  orientation = "io",
+                                     rts = "vrs")
R> plot(kaoliubccLeon)
\end{verbatim}

\begin{figure}
\includegraphics[width=\linewidth]{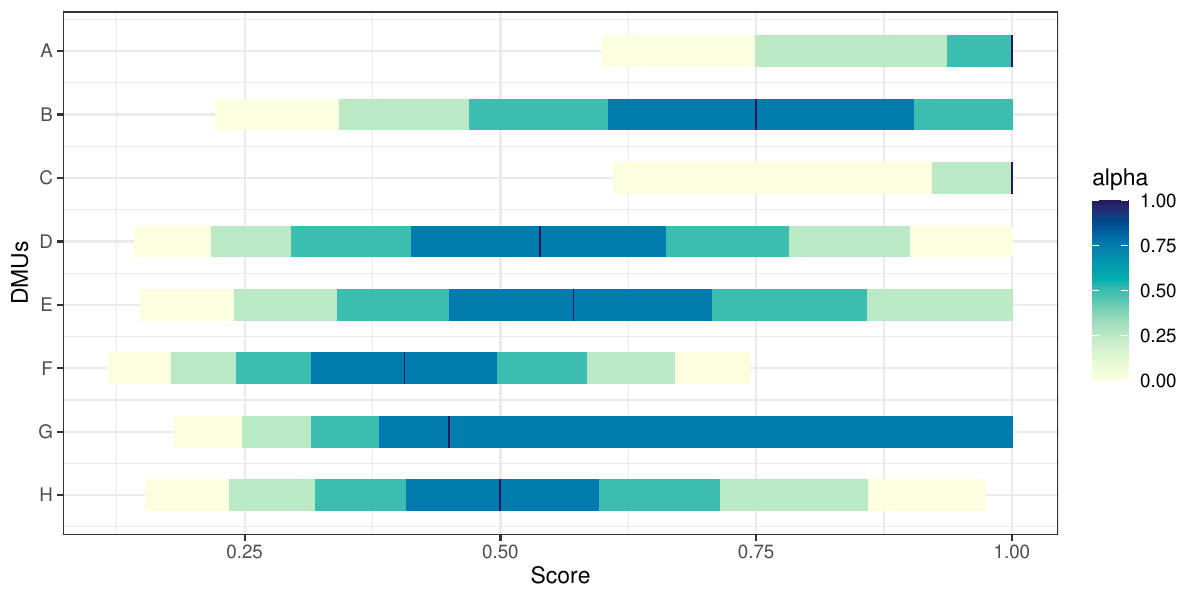}
\caption{Fuzzy efficiency scores obtained with Kao-Liu model. For each DMU, the fuzzy efficiency is represented by a coloured bar, in which the colour represents the membership degree $\alpha$ of the efficiency score.}\label{fig:kaoliu}
\end{figure}

The results are depicted in Figure~\ref{fig:kaoliu}. There, different types of DMUs can be found. For instance, F and H are completely inefficient, because no $\alpha$-cut contains the value 1 for the efficiency. B, D, and E are efficient only for some $\alpha$, but not for all, because 1 is contained in the $\alpha$-cuts only for sufficiently small values of $\alpha$. Finally, A, C, and G are efficient for all possible values of $\alpha$, since 1 is contained in all $\alpha$-cuts. However it is noteworthy to point out that A and C are ``crisp-efficient'' for large enough values of $\alpha$, in the sense that the $\alpha$-cut intervals are reduced to $\{1\}$ from a certain value of $\alpha$, while G always presents uncertainty. 

\subsection{Guo-Tanaka models}

Fuzzy models for symmetric triangular data under constant returns to scale were proposed by \citet{Guo2001}. These models are implemented in \texttt{modelfuzzy\_guotanaka}, specifically the input and output-oriented versions of the model in \citet[Equation~(16)]{Guo2001}. The fuzzy efficiencies are calculated using \citet[Equation~(17)]{Guo2001}.
According to their notation, the $\alpha$-cuts are called $h$-levels, and the (crisp) relative efficiencies and multipliers for the level $h = 1$ are obtained from the multiplier model (\texttt{model\_multiplier}). It is important to remark that the optimal solutions of the Guo-Tanaka models are not unique in general.

\begin{figure}
\includegraphics[width=\linewidth]{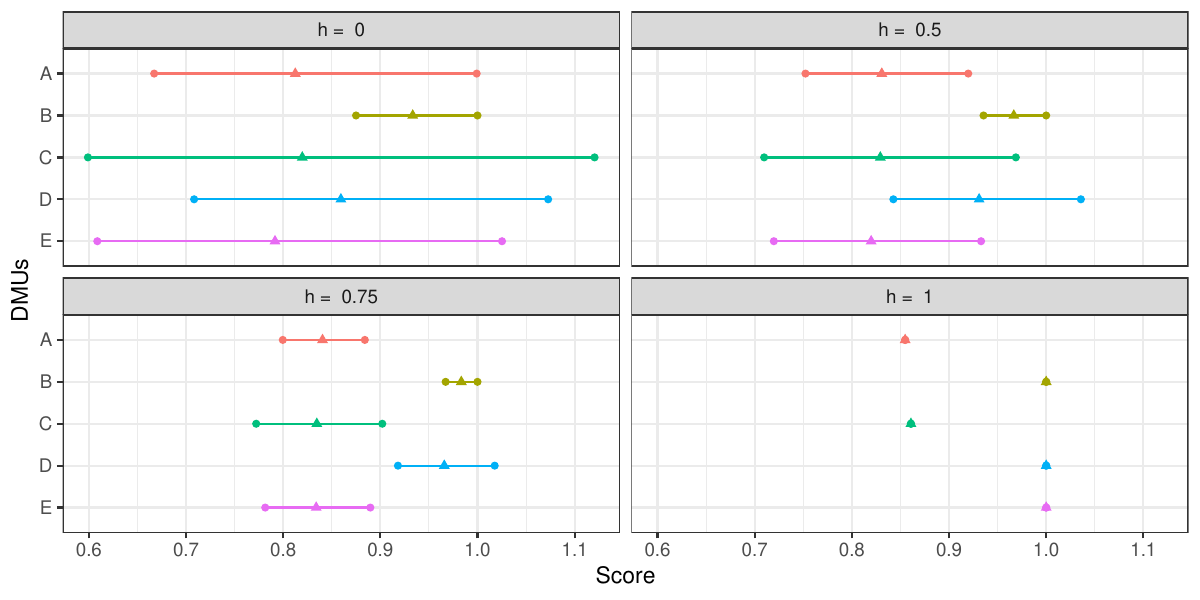}
\caption{Fuzzy efficiency scores obtained with Guo-Tanaka model. For each DMU and each $h$-level, the fuzzy efficiency is shown by a coloured line with three dots representing a triangular fuzzy number.}\label{fig:guo}
\end{figure}

We can replicate the results in \citet[p.~159]{Guo2001}:
\begin{verbatim}
R> dataGuo <- make_deadata_fuzzy(Guo_Tanaka_2001,
+                                inputs.mL = 2:3, inputs.dL = 4:5, 
+                                outputs.mL = 6:7, outputs.dL = 8:9)
R> guotanakaGuo <- modelfuzzy_guotanaka(dataGuo, h = c(0, 0.5, 0.75, 1))
R> plot(guotanakaGuo)
\end{verbatim}
The resulting scores are (non-symmetric) triangular fuzzy numbers for each $h$-level and can be extracted as usual with function \texttt{efficiencies} and plotted with \texttt{plot} (see Figure \ref{fig:guo}).

On the other hand, the \texttt{dea\_fuzzy} object returned by the model stores the results in the field \texttt{hlevel}, whose structure is similar to the structure of \texttt{alphacut} shown in Figure~\ref{fig:alphatree}. The difference is that there are not \texttt{Worst} and \texttt{Best} fields, and hence, the fields \texttt{DMU$_1$}, \texttt{DMU$_2$}, $\ldots$ (in which there are stored the results about efficiencies and multipliers) hang directly from the field \texttt{DMU}.

As a complement to Guo-Tanaka models, we have implemented cross-efficiency fuzzy models (arbitrary, aggressive and benevolent formulations) in function \texttt{cross\_efficiency\_fuzzy} for \texttt{dea\_fuzzy} objects returned by \texttt{modelfuzzy\_guotanaka}. For example,
\begin{verbatim}
R> crossGuo <- cross_efficiency_fuzzy(guotanakaGuo) 
\end{verbatim}
Alternatively, we can execute a Guo-Tanaka model internally, producing the same result:
\begin{verbatim}
R> crossGuo <- cross_efficiency_fuzzy(dataGuo, h = c(0, 0.5, 0.75, 1)) 
\end{verbatim} 

\subsection{Possibilistic models}

\begin{figure}
\includegraphics[width=\linewidth]{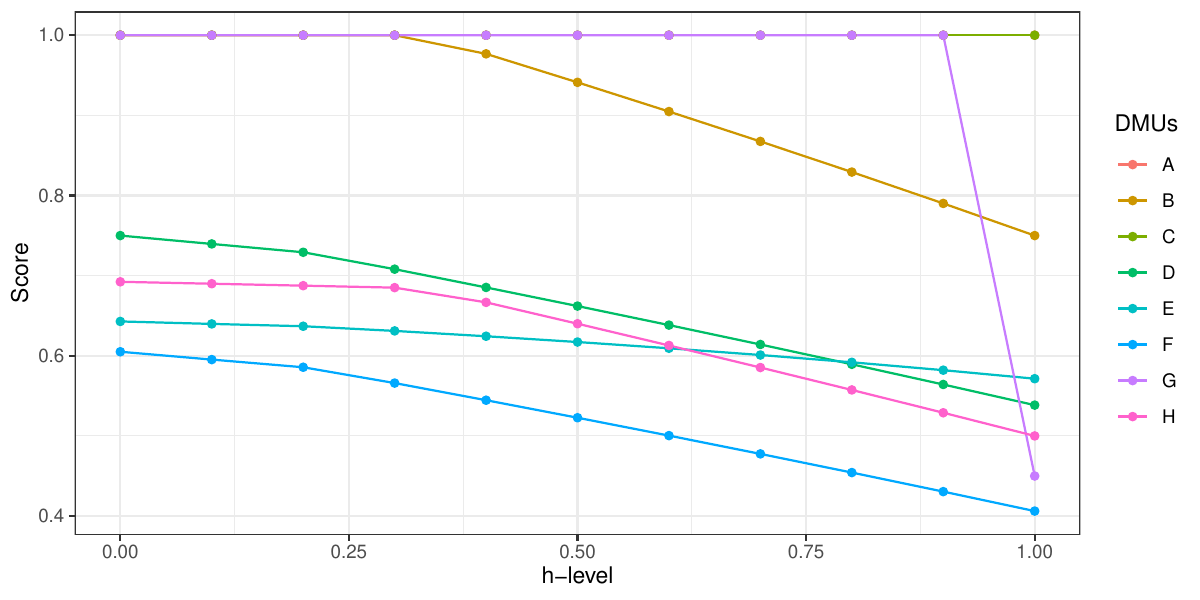}
\caption{Efficiency scores obtained with possibilistic model. For each DMU and each $h$-level, the efficiency is a crisp number represented by a coloured dot.}\label{fig:poss}
\end{figure}

Possibilistic fuzzy DEA models proposed by \citet{Leon2003} represent a generalization of the basic radial models to the fuzzy framework. By means of \texttt{modelfuzzy\_possibilistic}, we can replicate the results in \citet[p.~416]{Leon2003}:
\begin{verbatim}
R> possLeon <- modelfuzzy_possibilistic(dataLeon, h = seq(0, 1, by = 0.1), 
+                                       orientation = "io", rts = "vrs")
R> plot(possLeon)
\end{verbatim}
Note that, as in Guo-Tanaka models, the $\alpha$-cuts are called $h$-levels but, in this case, efficiency scores are crisp numbers for each $h$-level. The results are stored in the field \texttt{hlevel} of the \texttt{dea\_fuzzy} object returned by the model, and can be extracted as usual with functions \texttt{efficiencies} and \texttt{lambdas}. Moreover, function \texttt{plot} also works (see Figure \ref{fig:poss}).

\section{Malmquist index}
\label{sec:malm}

Classical DEA models provide a still picture of the performance of the DMUs. However, the activities of the DMUs often vary over time and therefore other types of models are needed in order to analyse said time variation. One of the most popular techniques is the Malmquist methodology, in which the DMUs are evaluated with respect to efficient frontiers corresponding to different time periods and different returns to scale.

Let $X^t,Y^t$ be the input and output data matrices, where $t=1,\ldots ,T$ denotes the time period. We define the CRS production possibility set at $t$ in the \emph{contemporary} (usual) way by $P^t=P\left( X^t,Y^t\right) $, according to \eqref{eq:p}. Analogously, we define the VRS production possibility set $P_B^t=P_B\left( X^t,Y^t\right) $ according to \eqref{eq:pb}.

Given an activity $\left( \mathbf{x},\mathbf{y} \right) \in \mathbb{R}_{>0}^{m+s}$, we define input and output-oriented ``distance'' functions at $t=1,\ldots,T$ by
\begin{equation}
\label{eq:DIDO}
\textrm{(a) } D_I^t\left( \mathbf{x},\mathbf{y} \right) =\inf \left\{ \theta \ \ | \ \ \left( \theta \mathbf{x},\mathbf{y} \right) \in P^t\right\},
\qquad
\textrm{(b) } D_O^t\left( \mathbf{x},\mathbf{y} \right) =\left( \sup \left\{ \eta \ \ | \ \ \left( \mathbf{x},\eta \mathbf{y} \right) \in P^t\right\}\right) ^{-1},
\end{equation}
respectively. Note that for a DMU at $t$, \eqref{eq:DIDO} (a) and (b) are the CCR efficiency score $\theta ^*$ and the inverse of $\eta ^*$ of the DMU (see \eqref{eq:ccr}), respectively. Moreover, we can define the VRS versions of \eqref{eq:DIDO}, $D_{IB}^t$ and $D_{OB}^t$ considering $P_B^t$ instead of $P^t$. In the following equations, $D$ can be $D_I$ or $D_O$, depending on the orientation.

According to \cite{Fare1994}, the \emph{Malmquist index} of DMU$_o$ at $t<T$ is given by
\begin{equation}
\label{eq:mifgnz}
MI_o^t=\left( \frac{D^t ( \mathbf{x}_o^{t+1},\mathbf{y}_o^{t+1}) \cdot D^{t+1} ( \mathbf{x}_o^{t+1},\mathbf{y}_o^{t+1}) }{D^t ( \mathbf{x}_o^{t},\mathbf{y}_o^{t}) \cdot D^{t+1} ( \mathbf{x}_o^{t},\mathbf{y}_o^{t}) }\right) ^{1/2} .
\end{equation}
The CRS decomposition of \eqref{eq:mifgnz} is given by
\begin{equation}
\label{eq:micrs}
MI_o^t = TC_o^t\cdot EC_o^t,
\end{equation}
with
\begin{equation}
\label{eq:tc}
TC_o^t=\left( \frac{D^t ( \mathbf{x}_o^{t+1},\mathbf{y}_o^{t+1})\cdot D^t ( \mathbf{x}_o^{t},\mathbf{y}_o^{t})}{D^{t+1} ( \mathbf{x}_o^{t+1},\mathbf{y}_o^{t+1})\cdot D^{t+1} ( \mathbf{x}_o^{t},\mathbf{y}_o^{t})}\right) ^{1/2} ,
\end{equation}
\begin{equation}
\label{eq:ec}
EC_o^t=\frac{D^{t+1} ( \mathbf{x}_o^{t+1},\mathbf{y}_o^{t+1})}{D^t ( \mathbf{x}_o^{t},\mathbf{y}_o^{t})},
\end{equation}
being the \emph{technical change} and the \emph{efficiency change} of DMU$_o$ at $t$, respectively. On the other hand, the VRS decomposition of \eqref{eq:mifgnz} is given by
\begin{equation}
\label{eq:tcpechsech}
MI_o^t = TC_o^t\cdot PECH_o^t\cdot SECH_o^t,
\end{equation}
with $TC_o^t$ given by \eqref{eq:tc} and
\begin{equation}
\label{eq:pech}
PECH_o^t=\frac{D_B^{t+1} ( \mathbf{x}_o^{t+1},\mathbf{y}_o^{t+1})}{D_B^{t} ( \mathbf{x}_o^{t},\mathbf{y}_o^{t})},
\end{equation}
\begin{equation}
\label{eq:sech}
SECH_o^t=\frac{D^{t+1} ( \mathbf{x}_o^{t+1},\mathbf{y}_o^{t+1})\cdot D_B^{t} ( \mathbf{x}_o^{t},\mathbf{y}_o^{t})}{D^{t} ( \mathbf{x}_o^{t},\mathbf{y}_o^{t})\cdot D_B^{t+1} ( \mathbf{x}_o^{t+1},\mathbf{y}_o^{t+1})},
\end{equation}
being the \emph{pure efficiency change} and the \emph{scale change} of DMU$_o$ at $t$, respectively.

According to \cite{Ray1997,Grifell1999}, the \emph{Malmquist index} of DMU$_o$ at $t<T$ is given by
\begin{equation}
\label{eq:mird}
MI_o^t=\frac{D^t ( \mathbf{x}_o^{t+1},\mathbf{y}_o^{t+1})}{D^t ( \mathbf{x}_o^{t},\mathbf{y}_o^{t})}.
\end{equation}
The VRS decomposition of \eqref{eq:mird} is given by \eqref{eq:tcpechsech}, with $PECH_o^t$ given by \eqref{eq:pech} and
\begin{equation}
\label{eq:tc2}
TC_o^t=\frac{D^{t}_B ( \mathbf{x}_o^{t+1},\mathbf{y}_o^{t+1})}{D^{t+1}_B ( \mathbf{x}_o^{t+1},\mathbf{y}_o^{t+1})},
\end{equation}
\begin{equation}
\label{eq:sech2}
SECH_o^t=\frac{D^{t} ( \mathbf{x}_o^{t+1},\mathbf{y}_o^{t+1})\cdot D_B^{t} ( \mathbf{x}_o^{t},\mathbf{y}_o^{t})}{D^{t} ( \mathbf{x}_o^{t},\mathbf{y}_o^{t})\cdot D_B^{t} ( \mathbf{x}_o^{t+1},\mathbf{y}_o^{t+1})}.
\end{equation}

According to \cite{Grifell1999}, a \emph{generalized} Malmquist index is defined by \eqref{eq:tcpechsech} with $TC_o^t$ given by \eqref{eq:tc2}, $PECH_o^t$ given by \eqref{eq:pech} and
\begin{equation}
\label{eq:sech3}
SECH_o^t=\frac{D^{t} ( \mathbf{x}_o^{t+1},\mathbf{y}_o^{t})\cdot D_B^{t} ( \mathbf{x}_o^{t},\mathbf{y}_o^{t})}{D^{t} ( \mathbf{x}_o^{t},\mathbf{y}_o^{t})\cdot D_B^{t} ( \mathbf{x}_o^{t+1},\mathbf{y}_o^{t})}.
\end{equation}

Moreover, according to \cite{Fare1997} a \emph{biased} Malmquist index can be computed from a \emph{biased} technical change, given by
\begin{equation}
\label{eq:maobib}
TC_o^t=MATECH_o^t\cdot OBTECH_o^t\cdot IBTECH_o^t,
\end{equation}
with
\begin{equation}
\label{eq:ma}
MATECH_o^t=\frac{D^{t} ( \mathbf{x}_o^{t},\mathbf{y}_o^{t})}{D^{t+1} ( \mathbf{x}_o^{t},\mathbf{y}_o^{t})},
\end{equation}
being the \emph{magnitude of technical change} of DMU$_o$ at $t$, and
\begin{equation}
\label{eq:obib1}
OBTECH_o^t=\left( \frac{D^{t} ( \mathbf{x}_o^{t+1},\mathbf{y}_o^{t+1})\cdot D^{t+1} ( \mathbf{x}_o^{t+1},\mathbf{y}_o^{t})}{D^{t+1} ( \mathbf{x}_o^{t+1},\mathbf{y}_o^{t+1})\cdot D^{t} ( \mathbf{x}_o^{t+1},\mathbf{y}_o^{t})}\right) ^{1/2},
\end{equation}
\begin{equation}
\label{eq:obib2}
IBTECH_o^t=\left( \frac{D^{t+1} ( \mathbf{x}_o^{t},\mathbf{y}_o^{t})\cdot D^{t} ( \mathbf{x}_o^{t+1},\mathbf{y}_o^{t})}{D^{t} ( \mathbf{x}_o^{t},\mathbf{y}_o^{t})\cdot D^{t+1} ( \mathbf{x}_o^{t+1},\mathbf{y}_o^{t})}\right) ^{1/2},
\end{equation}
being the \emph{output} and \emph{input bias indices} of DMU$_o$ at $t$, respectively. Different returns to scale can be considered in the computation of the biased technical change, using $D$ (CRS) or $D_B$ (VRS) in \eqref{eq:ma}, \eqref{eq:obib1} and \eqref{eq:obib2}.

All these indices can be computed in a \emph{sequential} way (instead of the usual \emph{contemporary} way) by considering the production possibility sets by $P^{\leq t}=P(X^{\leq t},Y^{\leq t})$ (CRS) and $P_B^{\leq t}=P_B(X^{\leq t},Y^{\leq t})$ (VRS), with $X^{\leq t},Y^{\leq t}$ being the input and output data matrices, respectively, of all the DMUs at time periods $\leq t$ \citep{Shestalova2003}. The new corresponding ``distance'' functions can be denoted by $D^{\leq t}$ and $D_B^{\leq t}$, and they must be used instead of $D^t$ and $D_B^t$.

Finally, the technical change can be defined in a \emph{global} way \citep{Pastor2005}. Its CRS version is given by
\begin{equation}
\label{eq:tcglobal}
TC_o^t=\frac{D^{t} ( \mathbf{x}_o^{t},\mathbf{y}_o^{t})\cdot D^{\leq T} ( \mathbf{x}_o^{t+1},\mathbf{y}_o^{t+1})}{D^{t+1} ( \mathbf{x}_o^{t+1},\mathbf{y}_o^{t+1})\cdot D^{\leq T} ( \mathbf{x}_o^{t},\mathbf{y}_o^{t})},
\end{equation}
where $D^{\leq T}$ denotes the ``distance'' function computed using a global production possibility set $P^{\leq T}$ considering all the DMUs of all time periods. A VRS version of \eqref{eq:tcglobal} is constructed by taking $D_B$ instead of $D$.

In order to read Malmquist datasets, the function \texttt{make\_malmquist} can deal with datasets in \emph{wide} or \emph{long} formats. \emph{Wide} format datasets are characterized by having the data for different time periods in different columns. In this case, parameter \texttt{nper} must contain the number of time periods and \texttt{arrangement} must be \texttt{"horizontal"} (by default). On the other hand, \emph{long} format datasets have the data for different time periods in the same column and hence, an extra column specifying to which time period the data belongs is required. In this case, parameter \texttt{percol} must contain the position of the column with the time periods and \texttt{arrangement} must be \texttt{"vertical"}. The rest of parameters are analogous to those in \texttt{make\_deadata} function. For example,
\begin{verbatim}
R> dataEconomy <- make_malmquist(Economy, ni = 2, no = 1,
+                                nper = 5, arrangement = "horizontal")
\end{verbatim}
for a wide format dataset, and
\begin{verbatim}
R> dataEconomyLong <- make_malmquist(EconomyLong, inputs = 3:4, outputs = 5,
+                                    percol = 2, arrangement = "vertical")
\end{verbatim}
for a long format dataset. In both cases, the result is a list with the different time periods. In turn, within each time period, the data is stored as a \texttt{deadata} object.

Once we have read a dataset, we compute Malmquist and other indices using function \texttt{malmquist\_index}. Parameter \texttt{datadealist} must contain the resulting list from function \texttt{make\_malmquist}. As usual, parameters \texttt{dmu\_eval} and \texttt{dmu\_ref} indicate which DMUs are evaluated and with respect to which DMUs they are evaluated, respectively. The orientation is given by \texttt{orientation}, that can take values \texttt{"io"} (input-oriented, by default) or \texttt{"oo"} (output-oriented). Parameter \texttt{rts} indicate which decomposition is applied to the Malmquist index: \eqref{eq:micrs} for \texttt{"crs"} (by default) or \eqref{eq:tcpechsech} for \texttt{"vrs"}.

Parameter \texttt{type1} determines the way in which we compute the production possibility sets: \texttt{"cont"} (contemporary, by default), \texttt{"seq"} (sequential) or \texttt{"glob"} (global). On the other hand, parameter \texttt{type2} determines the definition of the indices: \texttt{"fgnz"}  \citep{Fare1994} (by default), \texttt{"rd"} \citep{Ray1997,Grifell1999}, \texttt{"gl"} (generalized) \citep{Grifell1999} or \texttt{"bias"} (biased) \citep{Fare1997}. Finally, \texttt{tc\_vrs} is a logical parameter (\texttt{FALSE} by default) indicating if the biased technical change given by \eqref{eq:maobib} is computed under VRS or not.

Table~\ref{tab:malmquist} shows expressions used for different parameters combinations, apart from \texttt{orien\-tation}. Expressions for \texttt{type1 = "seq"} are computed considering sequential production possibility sets ($P^{\leq t},P_B^{\leq t},\ldots$) instead of contemporary ones ($P^t,P_B^t,\ldots$). Expressions \eqref{eq:ma}, \eqref{eq:obib1}, \eqref{eq:obib2} and \eqref{eq:tcglobal} can be computed under CRS or VRS. In the case \texttt{type1 = "cont"} or \texttt{"seq"}, \texttt{type2 = "bias"} and \texttt{rts = "vrs"}, expressions \eqref{eq:ma} \eqref{eq:obib1} and \eqref{eq:obib2} are computed under CRS if \texttt{tc\_vrs = FALSE} (by default) or under VRS if \texttt{tc\_vrs = TRUE}. Moreover, if \texttt{type1 = "glob"} then \texttt{type2} is irrelevant.

\begin{table}[ht]
\small
\label{tab:malmquist}
\centering
\begin{tabular}{|l|l|l|l|}
\hline
\texttt{type1} & \texttt{type2} & \texttt{rts} & Expressions \\ \hline
\multirow{6}{*}{\begin{tabular}[c]{@{}l@{}}\texttt{cont}\\
\texttt{seq}\end{tabular}} & \multirow{2}{*}{\texttt{fgnz}} & \texttt{crs} &
\eqref{eq:micrs}, \eqref{eq:tc}, \eqref{eq:ec} \\
\cline{3-4} & & \texttt{vrs} & \eqref{eq:tcpechsech}, \eqref{eq:tc}, \eqref{eq:pech}, \eqref{eq:sech} \\
\cline{2-4} & \texttt{rd} & \texttt{vrs} & \eqref{eq:tcpechsech}, \eqref{eq:pech}, \eqref{eq:tc2}, \eqref{eq:sech2} \\
\cline{2-4} & \texttt{gl} & \texttt{vrs} & \eqref{eq:tcpechsech}, \eqref{eq:pech}, \eqref{eq:tc2}, \eqref{eq:sech3}\\
\cline{2-4} & \multirow{2}{*}{\texttt{bias}} & \texttt{crs} & \eqref{eq:micrs}, \eqref{eq:ec}, \eqref{eq:maobib}, \eqref{eq:ma} (CRS), \eqref{eq:obib1} (CRS), \eqref{eq:obib2} (CRS) \\
\cline{3-4} & & \texttt{vrs} & \eqref{eq:tcpechsech}, \eqref{eq:pech}, \eqref{eq:sech}, \eqref{eq:maobib}, \eqref{eq:ma}, \eqref{eq:obib1}, \eqref{eq:obib2} \\ \hline
\multirow{2}{*}{\texttt{glob}} & \multicolumn{1}{c|}{\multirow{2}{*}{-}} & \texttt{crs} & \eqref{eq:micrs}, \eqref{eq:ec}, \eqref{eq:tcglobal} (CRS) \\
\cline{3-4} & \multicolumn{1}{c|}{} & \texttt{vrs} & \eqref{eq:tcpechsech}, \eqref{eq:pech}, \eqref{eq:sech2}, \eqref{eq:tcglobal} (VRS) \\ \hline
\end{tabular}
\caption{Expressions used for different parameters combinations in \texttt{malmquist\_index} function.}
\end{table}

Apart from the values of parameters \texttt{datadealist}, \texttt{dmu\_eval}, \texttt{dmu\_ref}, \texttt{orientation}, \texttt{rts}, \texttt{type1}, \texttt{type2} and \texttt{tc\_vrs}, the list returned by \texttt{malmquist\_index} contains all the indices involved in the computations: \texttt{mi}, \texttt{ec}, \texttt{tc}, \texttt{pech}, \texttt{sech}, \texttt{obtech}, \texttt{ibtech} and/or \texttt{matech}. Moreover, all the efficiency scores involved are stored in the field \texttt{eff\_all}, so users can build their own indices:
\begin{itemize}
\item \texttt{efficiency.*}: $D^t(\mathbf{x}_o^{t},\mathbf{y}_o^{t})$,
\item \texttt{efficiency\_t\_t1.*}: $D^t(\mathbf{x}_o^{t+1},\mathbf{y}_o^{t+1})$, 
\item \texttt{efficiency\_t1\_t.*}: $D^{t+1}(\mathbf{x}_o^{t},\mathbf{y}_o^{t})$,
\item \texttt{efficiency\_t\_xt1.*}: $D^t( \mathbf{x}_o^{t+1},\mathbf{y}_o^{t})$,
\item \texttt{efficiency\_t1\_xt1.*}: $D^{t+1}( \mathbf{x}_o^{t+1},\mathbf{y}_o^{t})$,
\item \texttt{efficiency.glob.*}: $D^{\leq T}(\mathbf{x}_o^{t},\mathbf{y}_o^{t})$,
\end{itemize}
where \texttt{*} can be \texttt{crs} or \texttt{vrs} (note that $D$ is replaced by $D_B$ in the \texttt{vrs} case). Moreover, if \texttt{type1 = "seq"} then $D^t$ and $D^{t+1}$ are replaced by $D^{\leq t}$ and $D^{\leq t+1}$, respectively.

We can replicate the results in \cite{Wang2011}:
\begin{verbatim}
R> malmquistEconomy <- malmquist_index(dataEconomy)
R> mi <- malmquistEconomy$mi
R> effch <- malmquistEconomy$ec
R> tech <- malmquistEconomy$tc
\end{verbatim}
Moreover, we can also replicate the results in \cite{Grifell1999}:
\begin{verbatim}
R> dataGrif <- make_malmquist(Grifell_Lovell_1999, percol = 1, dmus = 2,
+                             inputs = 3, outputs = 4,
+                             arrangement = "vertical")
R> fgnzGrif <- malmquist_index(dataGrif, orientation = "oo", rts = "vrs",
+                              type1 = "cont", type2 = "fgnz")
R> mi_fgnz <- fgnzGrif$mi 
R> rdGrif <- malmquist_index(dataGrif, orientation = "oo",
+                            type1 = "cont", type2 = "rd")
R> mi_rd <- rdGrif$mi
R> glGrif <- malmquist_index(dataGrif, orientation = "oo", rts = "vrs",
+                            type1 = "cont", type2 = "gl")
R> mi_gl <- glGrif$mi                     
\end{verbatim}

\section{Bootstrapping}
\label{sec:boots}

The bootstrap sampling method allows us to analyse the sensitivity of efficiency scores relative to variations in the efficient frontier.
We have implemented the bootstrapping algorithm proposed by \cite{Simar1998} for basic radial models in function \texttt{bootstrap\_basic}. Parameters \texttt{datadea}, \texttt{orientation}, \texttt{rts}, \texttt{L} and \texttt{U} acts as usual. On the other hand, parameter \texttt{B} indicates the number of bootstrap iterations ($2000$ by default) and \texttt{alpha} is a value between $0$ and $1$ ($0.05$ by default) determining the confidence intervals. Moreover, parameter \texttt{h} represents the bandwidth of smoothing windows. By default, \texttt{h} $ = 0.014$ but the optimal bandwidth factor can also be calculated following the proposals of \cite{Silverman1998} and \cite{Daraio2007}. So, \texttt{h = "h1"} is the optimal bandwidth referred to as ``robust normal-reference rule'' \citep[p.~60]{Daraio2007}, \texttt{h = "h2"} is the value of \texttt{h1} but with the factor $0.9$ instead of $1.06$, \texttt{h = "h3"} is the value of \texttt{h1} adjusted for scale and sample size \citep[p.~61]{Daraio2007}, and \texttt{h = "h4"} is the bandwidth provided by a Gaussian kernel density estimate.

The result is a list with the following fields of interest:
\begin{itemize}
\item \texttt{score}: efficiency scores from the corresponding basic radial model.
\item \texttt{score\_bc}: bias-corrected estimator of scores.
\item \texttt{bias}: bias of the score estimator.
\item \texttt{descriptives}: \texttt{mean\_estimates\_boot}, \texttt{var\_estimates\_boot} and \texttt{median\_estimates\-\_boot} contains the mean, variance and median of all the bootstrap iterations, respectively.
\item \texttt{CI}: confidence intervals of scores.
\item \texttt{estimates\_bootstrap}: results from each bootstrap iteration.
\end{itemize}

We can replicate the results in \citet[p.~58]{Simar1998}, but with $100$ iterations:

\begin{verbatim}
R> dataElectric <- make_deadata(Electric_plants, ni = 3, no = 1)
R> bootstrapElectric <- bootstrap_basic(dataElectric, rts = "vrs", B = 100)
R> head(bootstrapElectric$score_bc)
\end{verbatim}
\begin{verbatim}
    Coffeen Grant Tower Gudsonville   Meredosia      Newton        Fisk 
  0.8524186   0.9457947   0.9555025   0.9182929   0.9382163   0.8928644
\end{verbatim}

\begin{verbatim}
R> head(bootstrapElectric$CI)
\end{verbatim}
\begin{verbatim}
               CI_low     CI_up
Coffeen     0.8185240 0.8676870
Grant Tower 0.8566894 0.9991003
Gudsonville 0.8729732 0.9991638
Meredosia   0.8922955 0.9293679
Newton      0.8698494 0.9992010
Fisk        0.8662135 0.9063833
\end{verbatim}

\section{Non-parametric metafrontier analysis}
\label{sec:metaf}

One of the grounding hypotheses in DEA is that the DMUs must be comparable, not only in the sense that all DMUs should use the same inputs to produce the same outputs, but also that their production processes should be defined under the same technology framework. However, there are cases in which the objective is precisely to compare the efficient frontiers of groups of DMUs operating under different technologies, and to determine the so-called ``technology gap'' between those frontiers and the theoretical potential frontier of all the DMUs as a whole, which is called the \emph{metafrontier}. 

There are several approaches to the metafrontier analysis. On one hand, the \emph{stochastic metafrontier} analysis \citep{Battese2002} assumes a parametric relationship between inputs and outputs. On the other hand, there is the \emph{non-parametric metafrontier} analysis, that can be either \emph{concave} \citep{Battese2004} or \emph{non-concave} \citep{Tiedemann2011}. 

\begin{figure}
\includegraphics[width = \linewidth]{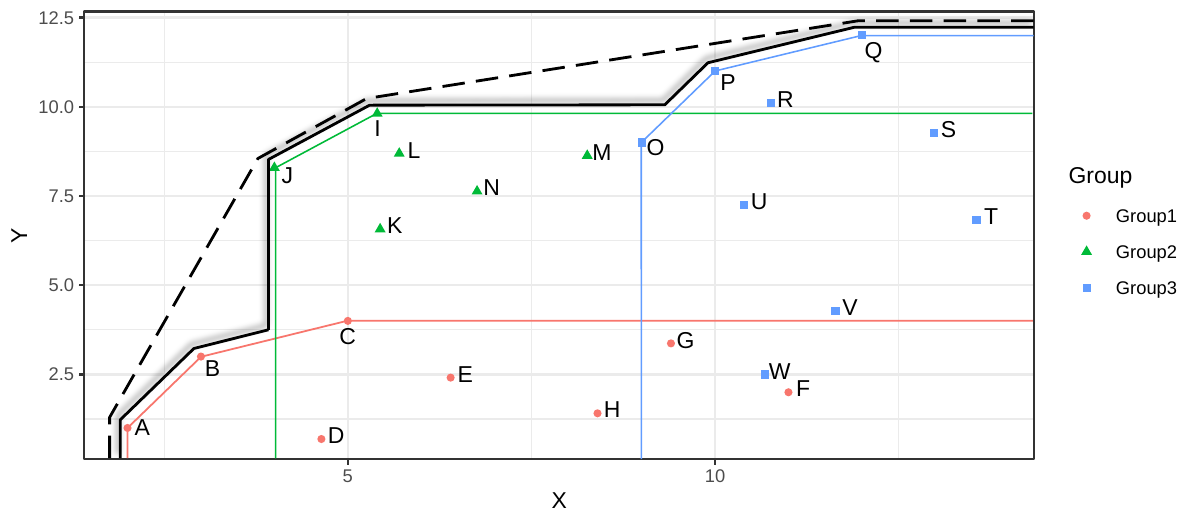}
\caption{Non-parametric metafrontier analysis. The efficient frontiers of each group (coloured lines) are represented together with the non-parametric concave (black dashed line) and non-concave (black shadowed line) metafrontiers. This figure has been created representing the inputs and outputs of the DMUs in a \texttt{XY}-plane.}\label{fig:metafr}
\end{figure}

In this section, we are going to illustrate with an example, how to compute the non-parametric metafrontiers using \textbf{deaR}. It is noteworthy that there is no specific function to do this, because it can be readily computed using the \texttt{dmu\_eval} and \texttt{dmu\_ref} parameters in the \texttt{model\_xxx} functions (see Section \ref{sec:run}). This example considers three groups of DMUs with one input \texttt{X} and one output \texttt{Y} operating under variable returns to scale (BCC model) with input orientation. Figure~\ref{fig:metafr} depicts the DMUs of this example, together with the three efficient frontiers and the resulting non-parametric metafrontiers.

First we create a synthetic dataset and we define the grouping:
\begin{verbatim}
R> DMUnames <- LETTERS[1:23]
R> X <- matrix(c(2, 3, 5, 4.64, 6.4, 11, 9.4, 8.4, 5.4, 4, 5.44, 5.7, 8.26,
+                6.76, 9, 10, 12, 10.76, 12.98, 13.56, 10.4, 11.64, 10.68),
+              nrow = 1, dimnames = list("X", DMUnames))
R> Y <- matrix(c(1, 3, 4, 0.69, 2.41, 2, 3.37, 1.41, 9.81, 8.29, 6.57, 8.69,
+                8.63, 7.63, 9, 11, 12, 9.73, 9.27, 6.83, 7.25, 4.27, 2.5),
+              nrow = 1, dimnames = list("Y", DMUnames))
R> datameta <- make_deadata(inputs = X, outputs = Y)
R> grouping <- list(G1 = 1:8, G2 = 9:14, G3 = 15:23)
\end{verbatim}

The following code makes use of the \texttt{dmu\_eval} and \texttt{dmu\_ref} parameters in order to evaluate the efficiency score of each DMU with respect to the three different efficient frontiers: 

\begin{verbatim}
R> eff <- sapply(grouping, function(x) sapply(grouping, function(y) {
+          model <- model_basic(datameta, dmu_eval = x, dmu_ref = y,
+                               rts = "vrs")
+          efficiencies(model) }))
\end{verbatim}

The efficiency of a DMU with respect to the non-parametric non-concave metafrontier is the minimum of the aforementioned efficiency scores evaluated for that DMU, after removing \texttt{NA}s for infeasible problems:

\begin{verbatim}
R> eff_meta <- sapply(eff,
+                     function(x) apply(x, MARGIN = 1, FUN = function(y)
+                                       min(y, na.rm = TRUE)))
R> show(eff_meta <- setNames(unlist(eff_meta), DMUnames))
\end{verbatim}
\begin{verbatim}
      A       B       C       D       E       F       G       H       I 
1.00000 1.00000 0.80000 0.43103 0.42266 0.22727 0.39787 0.26250 1.00000 
      J       K       L       M       N       O       P       Q       R 
1.00000 0.73529 0.76639 0.52217 0.59172 0.51711 1.00000 1.00000 0.49501 
      S       T       U       V       W 
0.37771 0.29499 0.38462 0.34364 0.25749 
\end{verbatim}

Finally, the efficiency with respect to the non-parametric concave metafrontier is the usual efficiency score considering all the DMUs as the evaluation reference set:

\begin{verbatim}
R> efficiencies(model_basic(datameta, rts = "vrs"))
\end{verbatim}
\begin{verbatim}
      A       B       C       D       E       F       G       H       I 
1.00000 0.84957 0.56461 0.43103 0.37294 0.20676 0.28194 0.25149 1.00000 
      J       K       L       M       N       O       P       Q       R 
1.00000 0.64855 0.76639 0.52217 0.56493 0.51711 0.89863 1.00000 0.49501 
      S       T       U       V       W 
0.37771 0.26545 0.35718 0.24889 0.22580 
\end{verbatim}

\section{Conclusions}
\label{sec:conc}

\textbf{deaR} package allows both researchers and practitioners to make use of a wide variety of DEA models. Among the conventional DEA models, the user can choose between radial and non-radial models (directional, additive, SBM, etc.), with different orientations and returns to scale, as well as consider variables with special features (non-controllable, non-discretionary or undesirable inputs/outputs). In addition, the package includes super-efficiency, cross-efficiency, Malmquist index and bootstrapping models. On the other hand, the versatility of \textbf{deaR} allows the user, for example, to calculate the generalized Farrell measure or to perform a metafrontier analysis.

A new feature unique to this package is the inclusion of several fuzzy DEA models. Currently, it contains the Kao-Liu, Guo-Tanaka and possibilistic models. Specifically, the Kao-Liu model can be applied to all conventional DEA models implemented in \textbf{deaR}, both efficiency and super-efficiency.

In an effort to improve the communication of the results, the package includes novel graphical representations such as the \textit{references graph} (see Figure \ref{fig:plotccr}), which aims to show the relationships among the DMUs that determine the efficient frontier and the inefficient DMUs. Also, in the case of the cross-efficiency analysis, the heat maps of the methods used are shown (see Figure \ref{fig:cross}). Moreover, the different fuzzy DEA models incorporate specific visualizations.

Finally, the architecture of the package makes it easy to upgrade with new features and models, e.g., stochastic or network DEA. Currently, we are also working on the design of new plots to aid the visualization of results, as we think this is a shortcoming in the DEA efficiency analysis. 

\bibliography{refs}
\bibliographystyle{abbrvnat}

\end{document}